\documentclass[useAMS,usenatbib,onecolumn]{mn2e}
\usepackage{graphicx}
\usepackage{bm}
\usepackage{epsfig}
\usepackage{amssymb, ulem}
\usepackage{amsmath}
\usepackage{empheq}
\usepackage{amsfonts}
\usepackage{cleveref}
\usepackage{bm}
\usepackage{booktabs}
\usepackage{tabularx}
\usepackage[makeroom]{cancel}
\newcommand{\bea}{\begin{eqnarray}}
\newcommand{\eea}{\end{eqnarray}}

\newcommand{\beq}{\begin{equation}}
\newcommand{\eeq}{\end{equation}}

\newcommand{\simless}[0]{\mathbin{\lower 3pt\hbox
   {$\rlap{\raise 5pt\hbox{$\char'074$}}\mathchar"7218$}}}
\newcommand{\simgreat}[0]{\mathbin{\lower 3pt\hbox
   {$\rlap{\raise 5pt\hbox{$\char'076$}}\mathchar"7218$}}}

\newcommand{\figref}[1]{figure \ref{#1}}

\newcommand{\capfigref}[1]{Figure \ref{#1}}

\newcommand{\eqnref}[1]{eq. (\ref{#1})} 
\newcommand{\eqnrefs}[1]{eqs. (\ref{#1})} 
\newcommand{\eqnrefbare}[1]{(\ref{#1})} 
\newcommand{\capeqnref}[1]{Equation (\ref{#1})}

\newcommand{\bfl}{\bmath \lambda}
\graphicspath{{}{figures/}}
\title[Density-velocity evolution from triaxial collapse]{Phase space dynamics of triaxial collapse:  Joint density-velocity evolution }
\author[Sharvari Nadkarni-Ghosh and Akshat Singhal]{Sharvari Nadkarni-Ghosh$^{1}$\thanks{E-mail:
sharvari@iitk.ac.in} and Akshat Singhal$^{2}$\thanks{E-mail:
akshat.singhal014@gmail.com} \\
$^{1}$Department of Physics, I.I.T. Kanpur, Kanpur, U.P. 208016 India \\
$^{2}$Department of Mathematics and Statistics, I.I.T. Kanpur, Kanpur, U.P. 208016 India}

\begin{document}
\date{}
\pagerange{\pageref{firstpage}--\pageref{lastpage}} \pubyear{}

\maketitle
\label{firstpage}

\begin{abstract}
We investigate the dynamics of triaxial collapse in terms of eigenvalues of the deformation tensor, the velocity derivative tensor and the gravity Hessian. Using the Bond-Myers model of ellipsoidal collapse, we derive a new set of equations for the nine eigenvalues and examine their dynamics in phase space. The main advantage of this form is that it eliminates the complicated elliptic integrals that appear in the axes evolution equations and is more natural way to understand the interplay between the perturbations. 

This paper focuses on the density-velocity dynamics. The Zeldovich approximation implies that the three tensors are proportional; the proportionality constant is set by demanding `no decaying modes'. We extend this condition into the non-linear regime and find that the eigenvalues of the gravity Hessian and the velocity derivative tensor are related as ${\tilde q}_d + {\tilde q}_v=1$, where the triaxiality parameter ${\tilde q} = (\lambda_{\mathrm{max}} - \lambda_{\mathrm{inter}})/(\lambda_{\mathrm{max}} - \lambda_{\mathrm{min}})$. This is a {\it new universal relation} holding true over all redshifts and a range of mass scales to within a few percent accuracy. The mean density-velocity divergence relation at late times is close to linear, indicating that the dynamics is dictated by collapse along the largest eigendirection. This relation has a scatter, which we show, is intimately connected to the velocity shear. Finally, as an application, we compute the PDFs of the two variables and compare with other forms in the literature. 
\end{abstract}

\maketitle
\section{Introduction}
Over the last decade or so, observations of the large scale structure in the universe have emerged as a very powerful probe to constrain cosmological parameters. The two main variables that characterize this structure are the fractional overdensity $\delta$ and the peculiar velocity ${\bf v}$. In the linear regime, the two observables are connected as $\nabla \cdot {\bf v} = - f H \delta$, where $H$ is the Hubble parameter and $f$ is the growth rate. $f$ is sensitive to the underlying cosmology and surveys such as 6dFGS \footnote{http://www.6dfgs.net/} or the future EUCLID \footnote{http://www.euclid-ec.org/} observe peculiar velocities either directly (e.g., \citealt{johnson_6df_2014}) or from redshift space distortions (e.g.,\citealt{majerotto_probing_2012}) with an aim to place precise constraints on $f$. It would be ideal if the data followed linear theory, but observations are sensitive to non-linear effects which can introduce a bias even on linear scales. Therefore, a theoretical understanding of the non-linear regime is imperative. Numerical simulations and perturbation theories are the two standard ways of tracking non-linear growth. However, both these methods have drawbacks. N-body codes are slow. Furthermore, they use a discrete representation of the density field and hence their results are shot-noise limited (for e.g., \citealt*{joyce_towards_2009}). Perturbation theories deal with smooth fields but they are not always guaranteed to converge and involved resummation techniques need to be invoked to get meaningful results (for e.g., \citealt*{matsubara_resumming_2008,matarrese_resumming_2007,nadkarni-ghosh_extending_2011,nadkarni-ghosh_modelling_2013}). Given the plethora of cosmological models, these features can prove to be restrictive. 

A third way to model the non-linear regime is to restrict the dynamics to simple geometries. Though based on local dynamics, such models often give theoretical insight into the underlying physics. The simplest among these is the spherical collapse model (spherical top-hat). It has been used in a myriad of ways starting from the mid-seventies to the present day. The critical density for collapse predicted by this model is an important ingredient in the mass function prescription given by \citet{press_formation_1974}. It has been widely used to understand the nature of non-linearities in a range of dark energy cosmologies from $\Lambda$CDM to early dark energy and quintessence models (for e.g., see \citealt{wintergerst_clarifying_2010} and references therein). \cite{kitaura_cosmological_2013} have used it in conjunction with Lagrangian perturbation theory to evolve perturbations through the shell-crossing regime. It has given valuable insights into the joint non-linear density-velocity evolution (\citealt{bilicki_velocity-density_2008,nadkarni-ghosh_non-linear_2013}). It has also been proposed as a control case to test the accuracy of N-body codes in the non-linear regime (\citealt{joyce_evolution_2012}). And last but not the least, it has been used to explain the famous NFW profile (\citealt*{navarro_structure_1996,lokas_universal_2000}) and as well as results of other simulations based on modified gravity (\citealt*{stabenau_n-body_2006,martino_spherical_2009}). 

Ellipsoidal or triaxial collapse is the next popular local model. It provides many improvements over the spherical geometry and has been in consideration for over five decades. Early studies (\citealt*{lynden-bell_large-scale_1964,lin_gravitational_1965}) examined the isolated ellipsoid in a non-expanding background. Cosmological extensions were performed by \citet{icke_formation_1973} and \citet{white_growth_1979} but under the assumption that the background did not exert external forces on the ellipsoid. \citet{bond_peak-patch_1996} included the effect of the background in terms of an external tidal field. \cite{nariai_dynamics_1972} and \cite{eisenstein_analytical_1995} provided a more complete analytic model that includes rotation as well. Since then, not much has changed in the theoretical ingredients of the model, however, the number of applications have been on the rise. Effects of non-radial motions on the growth rate and the resultant modifications to the Press-Schecter mass function were studied by several authors (for e.g., \citealt*{monaco_mass_1995, del_popolo_non_2000,del_popolo_ellipsoidal_2001,sheth_ellipsoidal_2001,kbt01}). Many authors developed alternatives to Press-Schecter that were based on statistics of collapse times derived from ellipsoidal collapse (\citealt*{audit_non-linear_1997,monaco_pinocchio_2002}). Others have obtained statistical measures of the non-linear density and velocity fields based on ellipsoidal collapse (\citealt*{fosalba_cosmological_1998-1,fosalba_cosmological_1998,scherrer_real-_2001,ohta_evolution_2003,ohta_cosmological_2004,lam_ellipsoidal_2008,lam_perturbation_2008}). \citet{angrick_triaxial_2010} used this model with additional components introduced to treat the virialization epoch. Very recently, \citet*{despali_ellipsoidal_2013} have advocated the use of ellipsoidal halo finders as an improvement over the spherical overdensity method to model shapes of haloes and this method has been applied to big numerical simulations to get insights into the shape distribution of dark matter haloes (\citealt{bonamigo_universality_2014}). Thus, spherical and ellipsoidal collapse models are not only used in isolation to get insights into the results of simulations, but they are also used in conjunction with numerical techniques to get semi-analytic estimates or for post-processing numerical data. 

One of the reasons why these simple geometries are so popular is that exact analytic solutions are available for homogenous perturbations evolving in pure matter cosmologies. The collapse in these cases is self-similar and the axes' lengths (or radius, in case of sphere) are the primary variables of interest. 
However, the main observationally relevant variables are the density and line of sight velocity, or more generally, the gravitational field and the peculiar velocity field. Thus it is more natural and interesting to directly understand how the these fields evolve in simple geometries. In case of spherical symmetry, the collapse is radial and only two variables suffice to describe the dynamics: density $\delta$ and the velocity divergence $\Theta$. In a recent paper \citet{nadkarni-ghosh_non-linear_2013}, hereafter N13, investigated the dynamics of these variables in a two-dimensional density-velocity divergence phase space \footnote{In N13, the velocity perturbation variable was $\theta = \Theta/3$}. A relation between the two variables was obtained by imposing the criterion of `no perturbations at the big bang time', and it was shown that this traces out a special curve in the 2D phase-space. The flow of perturbations is such that all perturbations, no matter where they start in phase space, eventually get attracted to this curve. Those that start along the curve, stay on it to a high degree of accuracy. The attracting nature established that this curve \footnote{In N13, this curve was termed the `Zeldovich curve' because it is the non-linear extension of the `no decaying modes' criterion, which is invoked in the linear Zeldovich approximation (\citealt{zeldovich_gravitational_1970}); here we call it the SC-DVDR} was the desired non-linear density-velocity relation and it was found that a combination of analytic forms given by \citet{bilicki_velocity-density_2008} and \citet{bernardeau_quasi-gaussian_1992} gave a good fit. 

In case of a triaxial ellipsoid, the situation is more involved. The full dynamics depends not only on the internal potential, but also on the external tidal field, which has been treated differently by different authors. The internal gravitational potential has a quadratic dependence on the length of the principle axes of the ellipsoid (\citealt{peebles80}). In this paper, we follow the model of \citet{bond_peak-patch_1996}, hereafter BM96, which assumes that the external tidal field is also along the principle axes of the ellipsoid throughout the evolution. The gravity field can then be described by three variables. These correspond to the eigenvalues of the tensor of second derivatives of the gravitational potential (henceforth called the `gravity Hessian'). In the absence of rotation, the velocity field also needs three variables. 
These are the eigenvalues of the tensor of partial derivatives of the velocity field (also sometimes called the `velocity deformation tensor' \footnote{In the past, these have also been referred to as the `gravitational shear' and `velocity shear' tensors respectively. However, we will keep this terminology only for the traceless part of these tensors.} and in this work referred to as the `velocity derivative tensor'). These six variables completely describe the gravitational dynamics of the ellipsoid. Since the potential of the system is intimately connected to its shape, these variables get connected to the eigenvalues of the deformation or displacement tensor.

In this work, we aim to study the joint evolution of density and velocity through the dynamics of these eigenvalues.
Starting from the BM96 set of equations for triaxial collapse, we obtain a set of coupled one-dimensional differential equations for the nine eigenvalues. This new set has no dependence on the elliptic functions that appear in the original equations for the axes lengths and is a well-defined dynamical flow. In \S\ref{sec:dynamics}, we derive these equations and by applying the `no perturbations at the big bang' condition, we get a relation between the eigenvalues of the gravity Hessian and the velocity derivative tensor. This traces out a subspace of perturbations, which we call the `Zeldovich subspace' and we find a new universal relation that describes the perturbations in this subspace. As a by-product, this analysis gives us insight into the relation between the density and the trace of the deformation tensor. In \S\ref{sec:2ddyn}, we examine the dynamics in the 2D $\delta-\Theta$ space. We investigate the scatter in the $\delta-\Theta$ relation and its relation to the velocity shear, thus providing a new insight based on local dynamics. Finally, as an application, in \S\ref{sec:application}, we numerically compute the marginal one-point probability distribution function (PDF) of the density and velocity divergence and compare them to some existing forms in the literature. We conclude in \S \ref{sec:conclusion}. Throughout this paper the terms `ellipsoidal' and `triaxial' are used interchangeably.

\section{Dynamics of the ellipsoid }
\label{sec:dynamics}

\subsection{Notation and equations}
Consider a uniform ellipsoidal distribution of cosmological fluid consisting of dark matter and dark energy evolving in a flat, homogenous and isotropic background. Let $\rho_{m,e}$ be the matter density inside the ellipsoid. It differs from the background matter density ${\bar \rho_m}$; the difference is characterized by the fractional density $\delta = \rho_{m,e}/{\bar \rho_m}-1$. The dark energy is the same inside and outside the ellipsoid and is described by a cosmological constant with density $\rho_\Lambda$. Let the origin be at the centre of the ellipsoid. The ellipsoid can be completely characterized by the evolution of its three principal axes. In this paper we follow the equations of BM96, which assume that the direction of the axes remains unchanged throughout the evolution. The physical coordinate of each axis $r_i$ can be written as $r_i = a_i(t) q$ ($i=1,2,3$), where $q$ is the comoving radius of the corresponding `Lagrangian sphere' (BM96): this is a sphere concentric with the ellipsoid whose mass equals that of the ellipsoid but whose density is the same as the background. $a_i$ are the `scale factors' of each axis and $a$ is the background scale factor. Throughout this paper we will use the subscript `$i$' to index the axes and `$init$' to denote initial conditions. Mass conservation during evolution implies $a^3 q^3 {\bar \rho_m} = a_1 a_2 a_3 q^3 \rho_{m,e}$ giving
\beq 
\delta = \frac{a^3}{a_1 a_2 a_3} -1.
\label{deltadef}
\eeq
The evolution of $a_i$ according to BM96 is \footnote{Our notation differs slightly from BM96: $\alpha_i \equiv b_i$ and $\lambda_{ext,i}\equiv \lambda_{ext,i}'$ in BM96. With this, the two terms $\frac{\delta}{3} + \frac{\delta b_i'}{2}$ in BM96 are equal to $\frac{\delta \alpha_i}{2}$ in our notation.}
 \beq 
 \frac{d^2 a_i}{dt^2} =  -4 \pi G \left[ \frac{{\bar \rho_m}}{3} -\frac{2}{3} \rho_{\Lambda}\right] a_i
  -4 \pi G \left[  {\bar \rho_m} \left\{  \frac{\delta \alpha_i}{2} + \lambda_{ext,i} \right\}\right] a_i,
\label{aieqn1}
\eeq
where 
\bea
\label{alphadef} 
\alpha_i &=& a_1a_2a_3 \int_0^\infty \frac{d \tau}{ (a_i^2 + \tau) \prod_{j=1}^{j=3}(a_j^2 + \tau)^{1/2}}  \;\; \; \; \mbox{with} \; \left(\sum_{i=1}^3 \alpha_i=2\right)\\
\lambda_{ext,i} &=& \frac{5}{4} \left(\alpha_i - \frac{2}{3} \right) \; \; \; \; \mbox{non-linear approx}. 
\label{lambdaext}
\eea
In \eqnref{aieqn1}, the first term in square brackets corresponds to the background potential and the second term to the perturbation potential.  The $\alpha_i$s are parameters that describe the internal potential of the ellipsoid and are computed using Carlson's elliptic integrals (\citealt{carlson87,carlson89,numrecipes}; see appendix \ref{derivdyn} for relations). $\lambda_{ext,i}(t)$ models the external tidal field. In this paper, we will use the non-linear approximation \footnote{A recent paper by \cite{angrick_triaxial_2010} experimented with using a hybrid model to compute halo mass functions. They used the non-linear expression of BM96 at early times and reverted back to the linear expression of BM96 after turn-around. They concluded that the hybrid model and the non-linear model gave approximately the same results. We prefer to use a single function than a piecewise one so we stick to the BM96 form}. 
Using standard definitions: ${\bar \rho_m} = \Omega_m \rho_c$, $\rho_\Lambda = \Omega_\Lambda \rho_c$ and $H^2 = 8 \pi G \rho_c/3$ and converting the time variable to $a$ gives 
\beq 
\frac{d^2 a_i}{d a^2} + \left\{\frac{1}{Ha} \frac{d}{d a} (H a)\right\} \cdot \frac{d a_i}{d a}  
= -\frac{3}{2 a^2}\left[\Omega_m(a) \left(\frac{1}{3} + \frac{\delta \alpha_i}{2} + \lambda_{ext,i} \right) - \frac{2}{3} \Omega_X(a) \right] a_i.
\label{aieqns2}
 \eeq
 Note that the $\Omega$ parameters are functions of $a$ and are related to their values today ($a=a_0$) by 
 \beq 
 \Omega_{m}(a)  = \frac{ \Omega_{m,0}  H_0^2a_0^3}{H^2 a^3};  \; \; \; \Omega_\Lambda(a) = \frac{ \Omega_{\Lambda,0}  H_0^2}{H^2 } .
 \label{omega}
 \eeq
 In this paper we will consider only two cosmologies. The EdS case with $\Omega_m=1$ and $\Omega_\Lambda =0$ and the $\Lambda$CDM case with $\Omega_m = 0.29$ and $\Omega_\Lambda = 0.71$.  
 The evolution of the ellipse is completely determined once six parameters are known: the three axes lengths $a_{i,init}$ and their velocities ${\dot a}_{i,init}$ at some initial epoch $a_{\mathrm{init}}$. 

An alternate description of the ellipse can be given by a set of nine {\it dimensionless} parameters 
\begin{subequations}
\label{params}
\begin{align} 
\label{ladef}
\lambda_{a,i} &= 1- \frac{a_i}{a} \;\;\;\;\; \\
\label{lvdef}
\lambda_{v,i} &= \frac{1}{H} \frac{{\dot a}_i}{a_i} -1 \\ 
\label{lddef}
\lambda_{d,i} &= \frac{\delta \alpha_i}{2} + \lambda_{ext,i}, 
 \end{align}
 \end{subequations}
 where $i=1,2,3$. The eigenvalues $\lambda_{d,i}$ are ordered as $\lambda_{d,1} \geq \lambda_{d,2} \geq \lambda_{d,3}$. This implies the ordering $\lambda_{v,1} \leq \lambda_{v,2} \leq \lambda_{v,3}$ and $\lambda_{a,1} \geq \lambda_{a,2} \geq \lambda_{a,3}$ at all times. 
  
The three $\lambda_{a}$ characterize the shape of the ellipse in terms of the deviation from the background. They correspond to the eigenvalues of the `comoving strain or deformation tensor' (see Appendix \ref{app:tensors} for definitions) . Because the ellipsoid is always deformed along its principle axes, the deformation tensor is diagonal at all times. When an axis is collapsing $\lambda_a \rightarrow 1$, whereas for an expanding axes, $\lambda_a \rightarrow -\infty$.

The three $\lambda_{v}$ capture the deviation of the velocity of each axes from the background Hubble flow. They correspond to the eigenvalues of the tensor of (scaled) velocity derivatives. The trace part gives the `expansion'  
\beq
\Theta = \frac{\nabla \cdot {\bf v}}{H} =\lambda_{v,1} + \lambda_{v,2} + \lambda_{v,3}.
\label{Thetadef}
\eeq  
where ${\bf v} = {\bf \dot r} - H {\bf r}$.  In our definition, a negative $\lambda_v$ implies a infall and a positive $\lambda_v$ implies expansion. The three eigenvalues completely describe the curl-free velocity field of this model. 

The three $\lambda_{d}$ correspond to the eigenvalues of the gravity Hessian (tensor of second derivatives of the gravitational field). \capeqnref{lddef} comprises of two terms; the first corresponds to the internal potential of the ellipse and the second to the external tidal field or the traceless gravitational tidal tensor. This definition and the fact that $\sum_i \alpha_i =2$, implies 
\beq 
\delta  = \lambda_{d,1} + \lambda_{d,2} + \lambda_{d,3}.
\label{delta_ld}
\eeq
This is also the consistency condition arising from Poisson's equation. 

Not all $\lambda$s are independent. \capeqnref{lddef} defines an implicit relation between the three $\lambda_{d}$ and the three $\lambda_a$ (this can be seen from the definitions of $\alpha_i$ and $\lambda_{a,i}$). In general, this relation is not linear. However, when the perturbations are small, the Zeldovich approximation implies $\lambda_{a,i}  = \lambda_{d,i} $ and $\lambda_{v,i} = -\lambda_{d,i}$. \capeqnref{delta_ld} reduces to 
\beq
\delta_{\mathrm{lin}} = \lambda_{a,1} + \lambda_{a,2} + \lambda_{a,3}. 
\label{deltalin}
\eeq
We emphasize that \eqnref{deltalin} is not valid in the non-linear regime and we will illustrate the difference in \S \ref{deltalinvsnl}. 
Usually it is standard to characterize the shape of the ellipse in terms of the `ellipticity' $e$ and `prolaticity' $p$ parameters (\citealt{bardeen_statistics_1986}). These are sometimes defined in terms of $\lambda_d$, but as has been emphasized in the literature (for e.g., \citealt{angrick_triaxial_2010}), such definitions are valid only in the linear regime. Connecting the shape to $\lambda_{d,i}$ in the non-linear regime necessarily involves solving for the three axes lengths. 

An important assumption in the framework of Bond \& Myers is that the principle axes of the deformation tensor and gravitational shear tensor coincide at all times during the evolution. If this assumption is violated then the ellipsoid can rotate and additional parameters need to be introduced to describe the dynamics.  A more general framework that allows for rotation has been introduced by \cite{eisenstein_analytical_1995}. 

In terms of the $\lambda$ parameters, the equations \eqnref{aieqns2} and its initial conditions are 
 \bea
 \label{aieqns3}
\frac{d^2 \bmath{a}}{d a^2} + \left\{\frac{1}{Ha} \frac{d}{d a} (H a)\right\} \cdot \frac{d {\bmath a}}{d a}  
&=& -\frac{3}{2 a^2}\left[\Omega_m(a) \left\{\frac{1}{3} + \bfl_{d}(a) \right\} - \frac{2}{3} \Omega_X(a) \right] {\bmath a}\\
 {\bmath a}_{\mathrm{init}} &=& a_{\mathrm{init}}(1-{\boldsymbol \lambda}_{a,\mathrm{init}}) \\
 \left. \frac{ d \bmath{a}}{d a} \right|_{a_{\mathrm{init}}} &=&  a_{\mathrm{init}} (1-{\boldsymbol \lambda}_{a,\mathrm{init}}) (1+{\boldsymbol \lambda}_{v,\mathrm{init}}).
 \eea
We have introduced boldface symbols for quantities that are 3-tuples. The product $(1-{\boldsymbol \lambda}_{a,\mathrm{init}}) (1-{\boldsymbol \lambda}_{v,\mathrm{init}})$ {\it does not} denote a dot product. It is  just the product of the corresponding components for each axes. The boldface ${\bf a}$ denotes the axes of the ellipse and $a$ denotes the scale factor of the background.

\subsection{Dynamics in phase space}
\subsubsection{Equations for the eigenvalues}
The nine (dimensionless) eigenvalues completely characterize the density, velocity and shape perturbations in this model of ellipsoidal collapse. In this paper, our focus is on the joint dynamics of density and velocity, hence, in principle, only six evolution equations are needed: one for each component of $\bfl_d$ and $\bfl_v$. These are obtained by their definitions in \eqnrefs{ladef}, \eqnrefbare{lvdef} and the evolution given by \eqnref{aieqns3}. However, it turns out that the equation for $\lambda_{d,i}$ involves complicated functions of $a_i$ which cannot be inverted easily since in the non-linear regime the relation between $\lambda_d$ and the axes is implicit (see Appendix \ref{derivdyn}). The system simplifies greatly if one adds the parameters $\lambda_a$ to the set. Thus, the net system for the nine eigenvalues is 
\begin{subequations}
\label{phspdyn}
\begin{align}
\label{dyn1}\frac{d \lambda_{a,i}}{d \ln a} &= -\lambda_{v,i}(1-\lambda_{a,i})\\
\label{dyn2}\frac{ d \lambda_{v,i}}{d \ln a}& = -\frac{1}{2} \left[ 3 \Omega_m(a) \lambda_{d,i}  -  \left\{ \Omega_m(a) - 2 \Omega_\Lambda(a)-2 \right \}  \lambda_{v,i}   + 2 \lambda_{v,i}^2\right] \\
\label{dyn3}\frac{d \lambda_{d,i}}{d \ln a} &= -(1+\delta) \left(\delta + \frac{5}{2} \right)^{-1} \left(\lambda_{d,i} + \frac{5}{6} \right)  \sum_{j=1}^3\lambda_{v,j} \\
\nonumber & + \left( \lambda_{d,i} + \frac{5}{6} \right) \sum_{i=1}^3 (1+\lambda_{v,i})  - \left( \delta + \frac{5}{2} \right)(1+\lambda_{v,i}) \\
\nonumber & + \sum_{j\neq i}\frac{ \left\{\lambda_{d,j} - \lambda_{d,i} \right\} \cdot \left\{(1-\lambda_{a,i})^2(1+\lambda_{v,i}) -(1-\lambda_{a,j})^2(1+\lambda_{v,j})\right\}}{(1-\lambda_{a,i})^2-(1-\lambda_{a,j})^2},
\end{align}
\end{subequations}
where
\beq
\nonumber \delta = \sum_{i=1}^3 \lambda_{d,i}.
\eeq
In general, the choice of $\lambda_{d,\mathrm{init}}$ (or alternatively $\lambda_{a,\mathrm{init}}$) and $\lambda_{v,\mathrm{init}}$ is independent. However, when the fluctuations are small, one generally employs the Zeldovich approximation \citep{zeldovich_gravitational_1970}, which states that the velocity is proportional to acceleration and the proportionality constant is set by requiring `no growing modes' (\citealt*{buchert_lagrangian_1992,susperregi_cosmic_1997}). Physically, this means that there are no perturbations at the big bang epoch. At early epochs, when $\Omega_m \approx 1$, this gives the relation $\bfl_a = \bfl_d$ and $\bfl_v = - \bfl_d$. More generally, $\bfl_v = -f(\Omega_m) \bfl_d$, where $f(\Omega_m) = \Omega_m^{0.55}$ is the linear growth rate (\citealt*{linder_cosmic_2005}; we have ignored the weak $\Lambda$ dependence). The linear density-velocity divergence is 
\beq \Theta_{\mathrm{lin}} = -f(\Omega_m) \delta_{\mathrm{lin}}.
\label{dvdrlin}
\eeq

\subsubsection{The Zeldovich subspace}

There are two approaches when we consider extension to the non-linear regime. One way is to initialize the system in the linear regime, evolve \cref{dyn1,dyn2,dyn3} into the non-linear regime and analyze the resulting $\bfl_d - \bfl_v$ at any later time. However, there is no guarantee that the resulting 6-tuples are universal; i.e. they may change with initial conditions. An alternate approach is to extend the essence of the Zeldovich approximation into the non-linear regime. That is, given three $\lambda_d$s one has to choose the three $\lambda_v$s (or vice versa) which satisfy the condition `no perturbations at the big bang'. This condition sets a special relation between the $\lambda_d$ and the $\lambda_v$ and we denote the resulting 6-tuples as   
$({\bfl}_d,\bfl_v^{Zel})$. This set defines a subspace in the 6D phase space which we call the `Zeldovich subspace'. Since there are only three independent parameters, this space is 3-dimensional. The superscript appears only on $\bfl_v$ to denote $\bfl_v$ is a function of $\bfl_d$ (our convention). Computationally, $\bfl_v^{Zel}$ is obtained as follows. Given the $\bfl_d$, first compute the $\bfl_{a}$ from the implicit relation \eqnref{lddef}. This acts as a known initial condition for \eqnref{aieqns3}. Then backward integrate \eqnref{aieqns3} with $\bfl_{v}$ as a unknown initial value which is solved for three simultaneous conditions ${\bf a}(a=0)=0$. It does not suffice to solve for only one or two of the axes. $\delta=0$ at $a=0$ only when all three axes are zero simultaneously. The value for $a_{\mathrm{init}}$ is taken to be the epoch for which the $\bfl_d - \bfl_v$ relation is desired. 

\begin{figure}
\includegraphics[width=16cm]{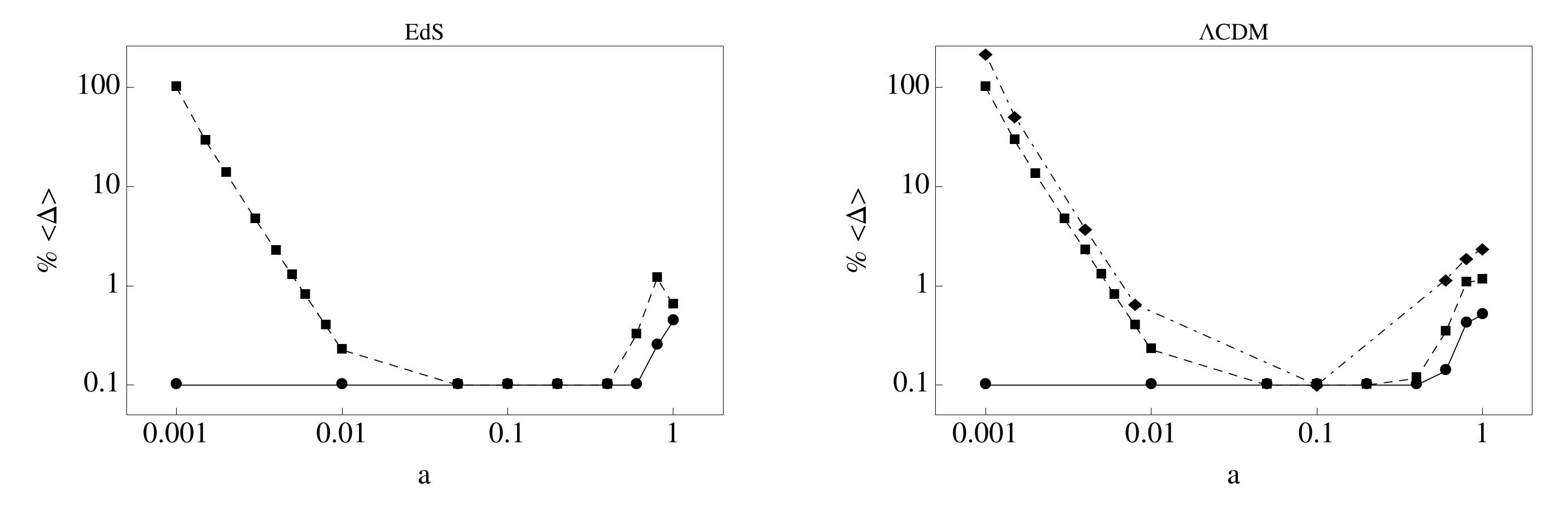}
\caption{The attracting nature of the Zeldovich subspace: the solid (dashed) lines correspond to $\bfl_{v,\mathrm{init}}=-\bfl_{d,\mathrm{init}}$ ($\bfl_{v,\mathrm{init}}=-2\bfl_{d,\mathrm{init}}$). The latter set has a 100\% deviation at the initial time $a=0.001$. For the $\Lambda$CDM case an additional non-collinear set $\bfl_{v,\mathrm{init}} = (-2,-3,-4) \cdot \bfl_{d,\mathrm{init}}$, where $\cdot$ implies component wise product) was considered (denoted by the dot-dashed lines). This set has a initial deviation of 216\%. The deviation decreases as a power law in $a$ scaling as $\sim a^{-2.6}$ for both EdS and $\Lambda$CDM cosmologies. }
\label{errfromzeld}
\end{figure}

The two approaches are complementary and it is not obvious that they will give identical results. A different set of initial conditions (for example, those that violated the Zeldovich approximation) could, in principle, generate future ($\bfl_d,\bfl_v$) values that need not satisfy the non-linear Zeldovich criterion (i.e., no perturbations at the big bang). However, we find that this is not the case. All perturbations, irrespective of whether they initially satisfy the Zeldovich approximation, satisfy the same $\bfl_d-\bfl_v$ relation at late times, suggesting a universal behaviour. This universality is tested as follows. This test is for universality is performed as follows. We track the trajectory of a set of initial points $(\bfl_{d,\mathrm{init}}, \bfl_{v,\mathrm{init}})$ using the eigenvalue equations and examine how much each trajectory deviates from the Zeldovich subspace at late times. The initial conditions at $a=0.001$ are drawn from the distribution given below (\cite{doroshkevich_spatial_1970}; see \citealt{rossi_initial_2012} and \citealt{angrick_ellipticity_2013} for recent extensions)
\beq 
 p(\lambda_{d,1}, \lambda_{d,2}, \lambda_{d,3})  = \frac{15^3}{8 \pi \sqrt{5} \sigma_G^6} \exp \left(-\frac{3I_1^2}{\sigma_G^2} + \frac{15 I_2}{2 \sigma_G^2}\right)\cdot (\lambda_{d,1}- \lambda_{d,2})(\lambda_{d,2} -\lambda_{d,3})(\lambda_{d,1} - \lambda_{d,3}) , 
\label{pdf}
\eeq
where $\sigma_G \equiv \sigma_G(R_f)$, the r.m.s. density fluctuation at the scale $R_f$, $I_1 = \lambda_{d,1} + \lambda_{d,2} + \lambda_{d,3}$ and $I_2= \lambda_{d,1}\lambda_{d,2} +  \lambda_{d,2}\lambda_{d,3}+  \lambda_{d,1}\lambda_{d,3}$. This PDF is gives the value at $a=1$; the value at $a=0.001$ is obtained by multiplying by the appropriate linear growth factor $D_+(a)= 5\Omega_{m,0}/2 \int_0^a [a'H(a')/H_0]^{-3} da$ (\citealt{dodelson}). Each point is evolved according to \cref{dyn1,dyn2,dyn3}. At a future epoch, a point on this trajectory is a 6-tuple denoted by $\{\bfl_d^{evol},{\bfl_v}^{evol}\}$. For this $\bfl_d^{evol}$, we compute the corresponding $\bfl_v^{Zel}$ as described above. This gives the corresponding point in the Zeldovich subspace denoted as $\{\bfl_d^{evol},\bfl_v^{Zel}\}$. The distance between $\{\bfl_d^{evol},{\bfl_v}^{evol}\}$ and $\{\bfl_d^{evol},\bfl_v^{Zel}\}$ is a measure of how close the trajectory gets to the Zeldovich subspace.
We define the relative deviation at any epoch as 
\beq 
\Delta (a) =  \frac{||{\boldsymbol \lambda}_v^{evol}(a) - {\boldsymbol \lambda}_v^{Zel}(a)||}{||{\boldsymbol \lambda}_v^{Zel}(a)||}, 
\label{deverr}
\eeq
where, $||{\bf x} - {\bf y}||$ denotes the norm $\sqrt{\sum_i (x_i - y_i)^2}$, $i=1,2,3$. 

\capfigref{errfromzeld} shows the average (over 50 points) relative deviation as a function of $a$ for the EdS (left panel) and $\Lambda$CDM models (right). For the EdS case, two sets of initial conditions were considered. The first set (solid line) was initialized at $a=0.001$ using the linear relation: $\bfl_{v,\mathrm{init}} =- \bfl_{d,\mathrm{init}}$ and the second (dashed line) with $\bfl_{v,\mathrm{init}} = -2 \bfl_{d,\mathrm{init}}$. In the first set, the error between the linear limit and the exact values that lie in the Zeldovich subspace was found to be 0.1\%. This value was set as a measure of the tolerance i.e. at any other epoch, the error was chosen to be the maximum of the mean deviation  and the tolerance. It was found that at most future epochs, the error stays within 0.1\% but rises to about 1\% near $a=1$.  The second set of initial conditions corresponds to 100\% deviation at $a=0.001$.  The relative deviation drops down exponentially from $\sim$ 100\% at $a=0.001$ to the sub percent level at $a=0.1$. This implies that even trajectories that start far off from the Zeldovich subspace eventually find their way onto it. For this set of initial conditions too, the error rises slightly near $a \sim1$. Similar behaviour is observed for the $\Lambda$CDM case. In this case, an additional third non-collinear set (dot-dashed line) of initial conditions was considered $\bfl_{v,\mathrm{init}} = \{-2,-3,-4\} \cdot \bfl_{d,\mathrm{init}}$ i.e., the x, y and z components are twice, thrice and four times the linear value. The initial deviation in this case is 216\%, but again it drops down exponentially. The latter two sets of initial conditions clearly illustrate the attracting nature of the Zeldovich subspace, but the reason for this late time rise is not yet fully clear. 

Some insight may be gained by looking at it from the point of view of Lagrangian Perturbation Theory (LPT). The Zeldovich approximation is the first order term in the LPT series \citep{buchert_lagrangian_1992} and it implies that there are no decaying modes at this order. In this construction, there may be decaying modes in the higher order 
solution. Requiring that there be `no perturbations at the big bang' is equivalent to demanding that there be `no growing modes' at every order in the LPT series (see \citealt{ehlers_newtonian_1997,nadkarni-ghosh_modelling_2013} for the general LPT series construction). Thus, the two solutions, one obtained by imposing `no growing modes' initially and then evolving and the other obtained by imposing `no growing modes at every order' are different and the differences could potentially grow at late times. Whether this explains the observed behaviour quantitatively is yet to be understood and is beyond the scope of this paper. Here, it suffices to note that compared to the initial deviation from the Zeldovich subspace ($\sim$ 100 \%) the final deviation is two orders of magnitude smaller.  

In recent work (N13), this technique was applied to spherical perturbations. It was shown that the non-linear density-velocity relation traced out a universal curve which satisfied the form 
\beq
\label{eqsphere}
\Theta_{\mathrm{sph}} = \left \{\begin{array}{cc}
\frac{3}{2} \Omega_m^{\gamma_1}\left[1-(1+\delta_{\mathrm{sph}})^{\frac{2}{3}\Omega_m^{\gamma_2}}\right] &-1\leq\delta_{\mathrm{sph}} <1 \\
&\\
\Omega_m^{\gamma_1 + \gamma_2}\left[(1+\delta_{\mathrm{sph}})^{\frac{1}{6}}-(1+\delta_{\mathrm{sph}})^{\frac{1}{2}} \right] & \delta_{\mathrm{sph}} \geq 1
\end{array}
\right.
\eeq
with $\gamma_1=0.56$ and $\gamma_2 =-0.01$ for a $\Lambda$CDM cosmology. This density-velocity divergence formula based on spherical collapse is a combination of the forms of \cite{bernardeau_quasi-gaussian_1992} and \cite{bilicki_velocity-density_2008} and henceforth will be referred to as the `SC-DVDR'. This curve is time-invariant for an EdS cosmology, but changes for the $\Lambda$CDM case due to the variation of $\Omega_m$ throughout evolution. 

\subsubsection{A universal non-linear $\bfl_d-\bfl_v$ relation}

\begin{figure}
\includegraphics[width=16cm]{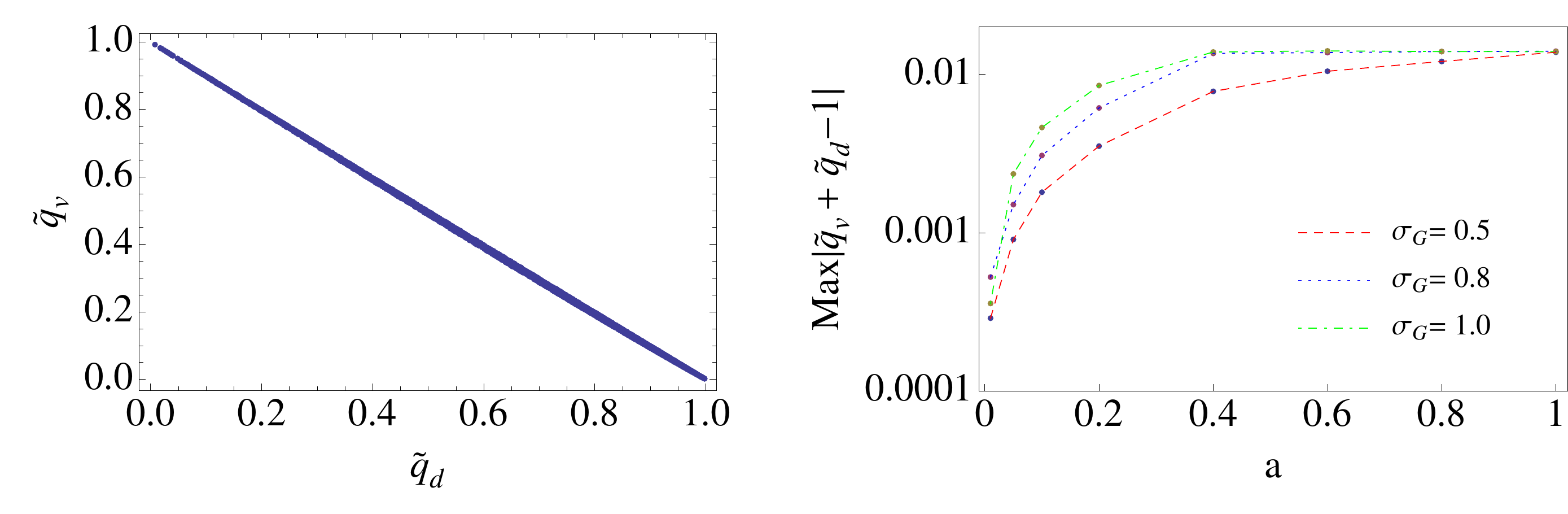}
\caption{A universal velocity-gravity relation: The left panel plots the triaxiality parameter paris $({\tilde q}_d, {\tilde q}_v)$ for a single realization ($\sigma_G=1$ at $a=1$) in the $\Lambda$CDM cosmology. The pairs lie on a straight line defined by ${\tilde q}_v + {\tilde q}_d = 1$. The error in the relation is plotted in the right panel. The error increases with epoch, but stays within 2\% for the range of $\sigma_G$s considered here.}
\label{ldlv}
\end{figure}
We find that it is possible to characterize the universality of the $\bfl_d - \bfl_v$ relation in terms of triaxiality parameters ($s$ and $q$). These are generally defined in the study of axis ratios \citep*{schneider_shapes_2012,nadkarni-ghosh_phase_2015}, in terms of the major and minor axes or alternatively in terms of the eigenvalues of the deformation tensor. Here, we generalize the definitions to eigenvalues of any 3-dimensional tensor.  
\bea
s &=&  \frac{1-\lambda_{\mathrm{max}}}{1-\lambda_{\mathrm{min}}}\\
q&=&\frac{1-\lambda_{\mathrm{inter}}}{1-\lambda_{\mathrm{min}}} \\
{\tilde q} &=& \frac{q-s}{1-s} = \frac{\lambda_{\mathrm{max}}- \lambda_{\mathrm{inter}}}{\lambda_{\mathrm{max}}-\lambda_{\mathrm{min}}},
\eea
where, $\lambda_{max/min/inter}$ denote the maximum, minimum and intermediate eigenvalues. 
By construction, 
${\tilde q}$ is smaller than unity. We find that, to a very good approximation, the $\bfl_d - \bfl_v$ relation is given by
\beq 
{\tilde q}_v + {\tilde q}_d = 1, 
\label{ldlveqn}
\eeq
where the subscripts $d$ and $v$ refer to the gravity Hessian and the velocity derivative tensor respectively. 
 It is easy to see that this relation is satisfied in the linear regime because $\lambda_{d, max} = -\lambda_{v,min}$ and vice versa. \capfigref{ldlv} illustrates this relation (left panel) and its accuracy (right panel). The left panel plots the $\{{\tilde q}_d,{\tilde q}_v\}$ pairs for a single realization ($\sim$ 10,000 points) corresponding to $\sigma_G =1$ at $a=1$ for the $\Lambda$CDM cosmology. As a measure of the accuracy of the relation, we compute $Max|{\tilde q}_v + {\tilde q}_d - 1|$, where the maximum is taken over 50,000 points (five realizations). The right panel plots this error as a function of $a$ for three different values of $\sigma_G$. The accuracy of the relation decreases with epoch, but stays within 2\%. This relation is a ratio of differences, and since the primary dependence of the velocity-gravity relation is through the linear growth factor, we expect this to be valid for other cosmologies as well (i.e. for other values of $\Omega_m$ and $\Omega_\Lambda$). Thus, it is universal, not only with respect to redshift and the mass scale ($\sigma_G$), but also with respect to certain standard dark energy models. 

 We found the above relation somewhat serendipitously while studying the behaviour of axis ratios \cite{nadkarni-ghosh_phase_2015}. The PDF of the ${\tilde q}$ parameter for the deformation tensor turns out to be an invariant of the dynamics (to percent level accuracy), and it was natural to investigate the behaviour of the corresponding ${\tilde q}$ parameter for the gravity Hessian and the velocity derivative tensor. The pattern of the PDFs for ${\tilde q}_v$ and ${\tilde q}_d$ suggested that ${\tilde q}_d + {\tilde q}_v$ was a constant, equal to one. The Zeldovich subspace is three dimensional. To completely describe this subspace analytically, one requires two additional independent equations connecting the $\lambda_d$s and $\lambda_v$s. More detailed investigations of the phase space equations will be necessary to construct appropriate invariants and are beyond the scope of this paper.

\subsubsection{2D $\lambda_d-\lambda_v$ subspaces}
\begin{figure}
\includegraphics[width=16cm]{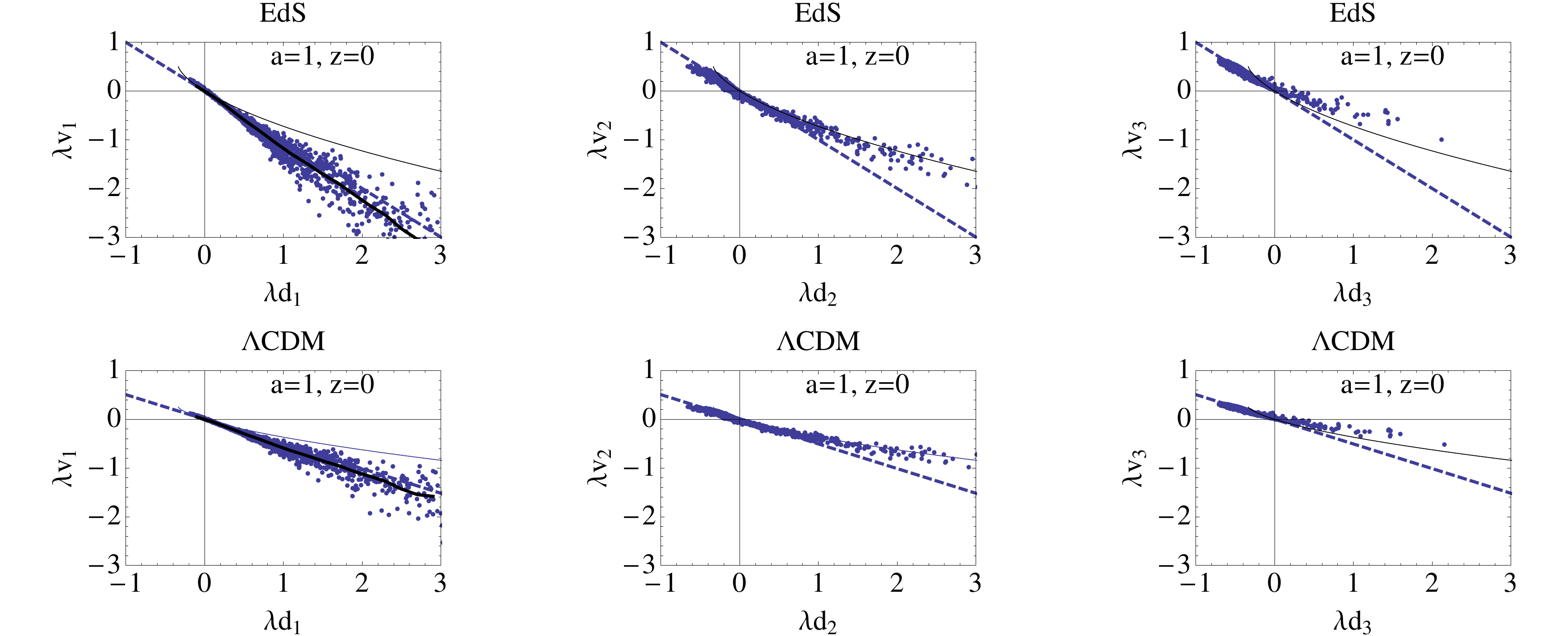}
\caption{Snapshot at $a=1$ of the 2D $\lambda_d-\lambda_v$ slices of the subspace. The solid line is the SC-DVDR given by \eqnref{eqsphere} and the dashed line is the linear relation $\lambda_v = -\lambda_d$. The first panel corresponds to the largest eigenvalues. The `mean' relation between them is very close to the linear relation. This is also responsible for the mean $\delta-\Theta$ relation being close to linear, although to a lesser extent (see \figref{deltatheta}). }
\label{lambdadandv}
\end{figure}
The entire Zeldovich subspace of the 6D phase space cannot be visualized directly. Nevertheless, it helps to visualize 2D subspaces defined by the parameters $(\lambda_{d,i}, \lambda_{v,i})$. \capfigref{lambdadandv} shows three such projections at $a=1$. 
In each plot, the dotted line shows the linear relation $\lambda_{v,i} = -f(\Omega_m)\lambda_{d,i}$. The solid curved line is the SC-DVDR given by \eqnref{eqsphere}; in this case, $\lambda_{d} = \frac{\delta_{\mathrm{sph}}}{3}$ and $\lambda_{v} = \frac{\Theta_{\mathrm{sph}}}{3}$, where 
$\delta_{\mathrm{sph}}$ and $\Theta_{\mathrm{sph}}$ are given by \eqnref{eqsphere}. The main interesting point observed here is that the `mean' relation between the largest eigenvalue $\lambda_{d,1}$ and the corresponding $\lambda_{v,1}$ is close to the linear one. It is clear that in the non-linear regime, each axes has a different relation between $\lambda_d$ and $\lambda_v$ i.e., the `vectors' $\bfl_d$ and $\bfl_v$ are no longer proportional. This means that although in the triaxial model, the gravitational shear tensor and velocity derivative tensor always have the same principle axes, their eigenvalues are not simple multiples of each other. This is related to the fact that the gravitational acceleration and peculiar velocity are not parallel to each other in the non-linear regime, which has been discussed earlier in the context of Lagrangian perturbation theory (for e.g., \citealt{bagla_new_1996,susperregi_cosmic_1997,nadkarni-ghosh_modelling_2013}). We also note the distinction between the breakdown of parallelism and breakdown of proportionality; for e.g. for the sphere with a radially dependent overdensity, the acceleration and velocity in the non-linear regime are always parallel at any point (both are radial), but yet as fields they are not proportional.

\subsection{$\delta$ vs. trace of the deformation tensor}
\label{deltalinvsnl}
\begin{figure}
\includegraphics[width=16cm]{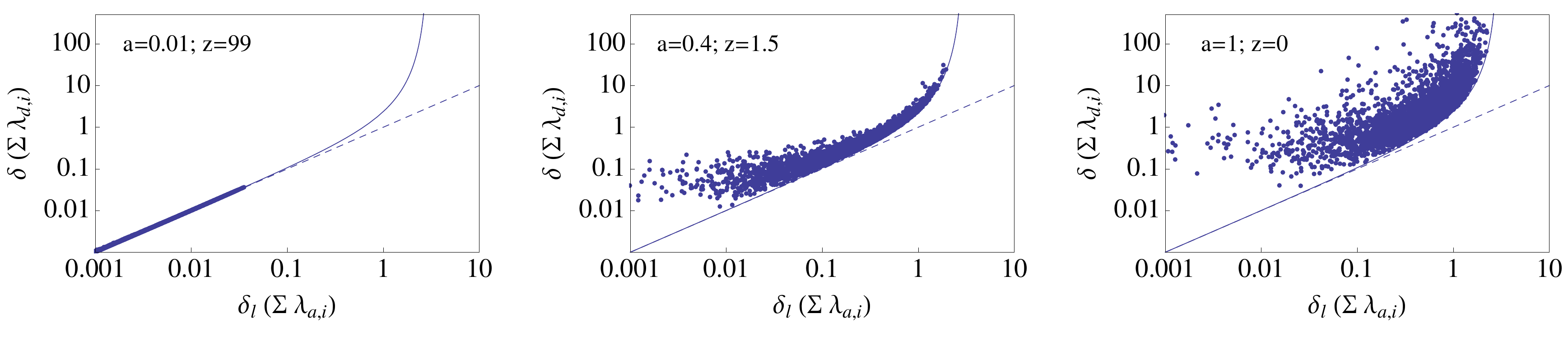}
\caption{Difference between the exact $\delta$ given by $\sum_i \lambda_{d,i}$ and the sum of the eigenvalues of the strain tensor which is used as an approximation. }
\label{devsa}
\end{figure}

Using the phase space evolution one can also investigate the relation between the exact non-linear density $\delta = \sum_i \lambda_{d,i}$ and its linear approximation $\delta_l = \sum_i \lambda_{a,i}$. \capfigref{devsa} shows the difference at three epochs: $a=0.01, 0.4$ and $1$. The dashed line denotes the linear relation and the solid line denotes the relation when all the axes are equal. In this case, $\lambda_a = \delta_{l,\mathrm{sph}}/3$, where $\delta_{l,\mathrm{sph}}$ is the linear spherical overdensity and from \eqnrefs{deltadef} and \eqnrefbare{ladef} we have 
\beq
\delta_{\mathrm{sph}} = \frac{1}{(1-\frac{\delta_{l,\mathrm{sph}}}{3})^3}-1.
\eeq
The dots indicate the numerically evolved data points. It is clear that at early times the linear approximation is valid, but at later epochs (middle and right panels), it fails for values as low as $\delta_l \sim 0.01$, which are small enough to be considered linear. For example, for some points with $\delta_l = 0.01$, the middle panel ($a=0.4$) shows the non-linear $\delta$ to be 10 times higher. The right panel ($a=1$) shows it to be 100 times higher. Thus, the linear approximation severely underestimates the actual value of $\delta$ and in particular ceases to be valid at late times even for linear densities. 

This discrepancy can be understood if one examines the mathematical expressions relating $\delta$ and $\delta_l$.  From \eqnrefs{deltadef} and \eqnrefbare{ladef}
\bea
\delta &=& \prod_{i=1}^3 (1-\lambda_{a,i})^{-1}  
\label{deltalareln}\\
&\approx & \delta_l + \sum_{i=1}^3 \lambda_{a,i}^2 +  \lambda_{a,1}\lambda_{a,2} +  \lambda_{a,2}\lambda_{a,3}  +  \lambda_{a,1}\lambda_{a,3} + \mathcal{O}(\lambda^3)
\eea
At early times, all $\lambda_a$s are small and the second order correction is small for all values of $\lambda_a$.  However, due to the asymmetry, it is possible to have two values of $\lambda_a$ large ($\sim \mathcal{O}(1)$) but with opposite signs. This asymmetry is already present in the initial conditions. The initial PDF given by \eqnref{pdf} has only 16\% probability of generating all $\lambda$s with the same sign (all positive or all negative). In 84\% of the cases, there are atleast two with different signs (this distribution stays more or less constant through the evolution; see Appendix \ref{app:percentages}). Suppose $\lambda_{a,1} = -\lambda_{a,3} \sim \mathcal{O}(1) $ and $\lambda_{a,2}$ is small. Then the second order term reduces to $\lambda_{a,1}^2 \sim \mathcal{O}(1)$, i.e. not small compared to the first order terms. To illustrate this, we consider one sample point in the right panel. This has values $\{\delta_l, \delta\} = \{0.004, 1.134\}$ and the individual $\bfl_a=\{0.708,-0.819,0.115\}$. The second order terms in this case add up to $0.5$, much larger than the linear value. Therefore, strictly speaking, the approximation that $\delta \sim \sum_{i=1}^3\lambda_{a,i}$ is valid only at very early epochs $a\lesssim 0.01$. 

In passing, we also note that all the data points lie above the solid curve which represents the spherical relation. This can be proven mathematically: for a fixed sum $\sum_i\lambda_{a,i}$, the l.h.s. of \eqnref{deltalareln} is minimum when all the axes are equal. 

\section{Dynamics in the $\delta -\Theta$ plane. }
\label{sec:2ddyn}
In this section, we analyse the dynamics in the 2D space corresponding to the variables $\delta = \sum \lambda_{d,i}$ and $\Theta=\sum \lambda_{v,i}$.  
\subsection{Ellipsoid vs. sphere}
\begin{figure}
\includegraphics[width=16cm]{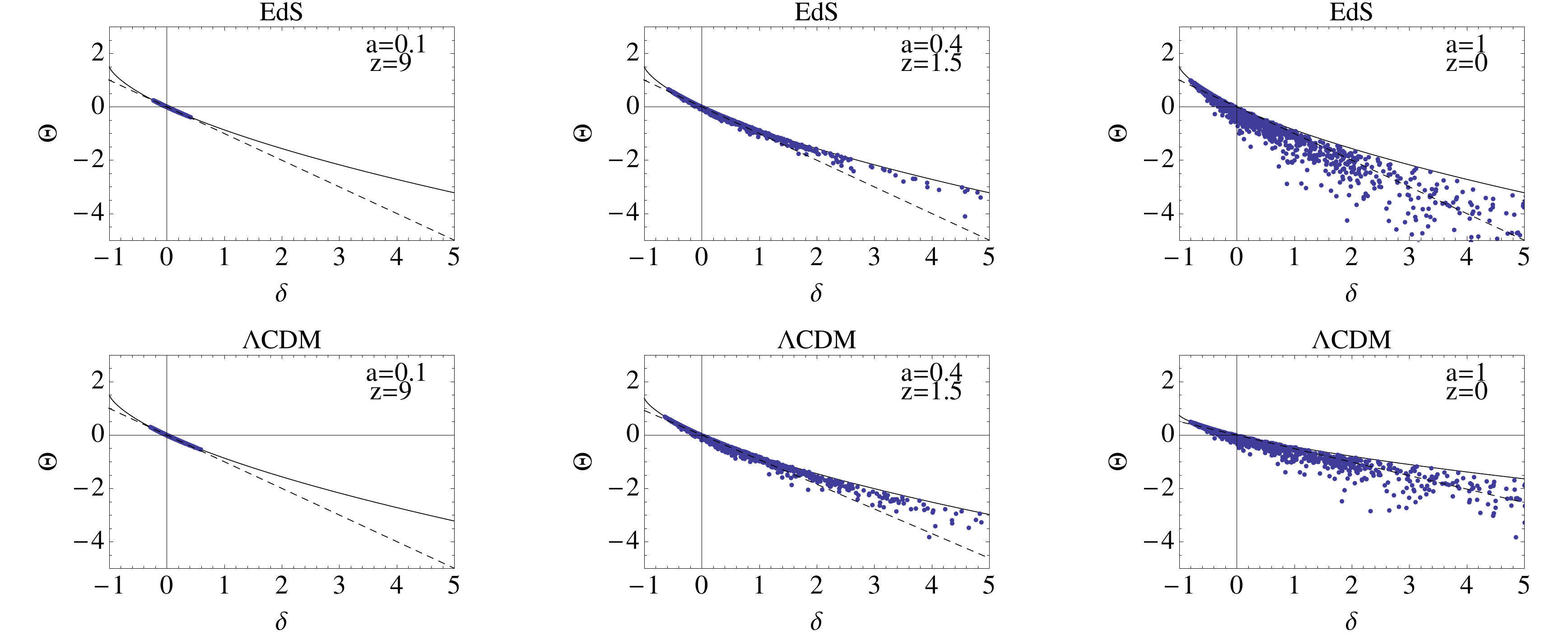}
\caption{The non-linear $\delta-\Theta$ relation based on triaxial dynamics. The dashed line is the linear relation and the solid line is the SC-DVDR given by \eqnref{eqsphere}. At early epochs, both the sphere and ellipsoid obey linear dynamics. Velocity shear effects induce a scatter for the triaxial dynamics. The average $\delta-\Theta$ relation is close to linear, though to a less extent than the $\lambda_{d,1}-\lambda_{v,1}$ relation (see \figref{lambdadandv}) because of the contribution from the second and third axes. }
\label{deltatheta}
\end{figure}

\begin{figure}
\includegraphics[width=16cm]{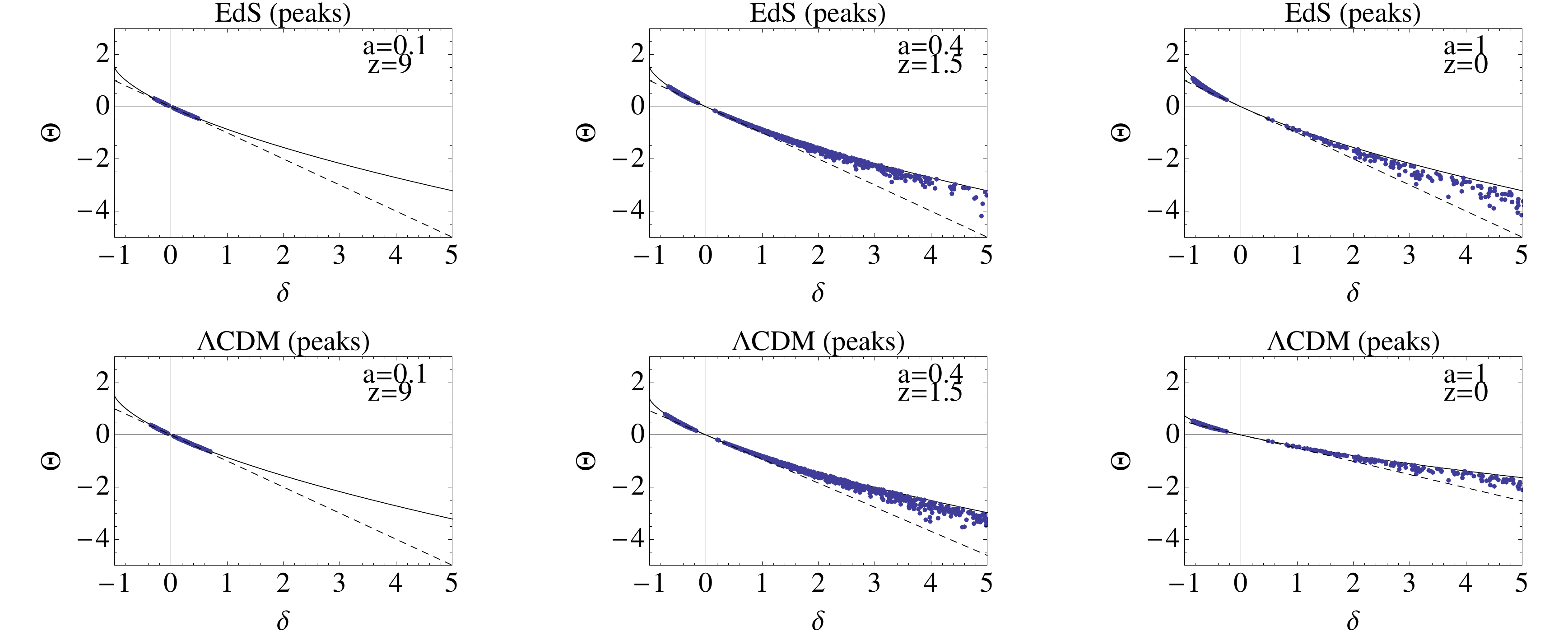}
\caption{The $\delta-\Theta$ relation for initial conditions such that all $\lambda$s have the same sign. The values lie much closer to the SC-DVDR than in the general case (\figref{deltatheta}).}
\label{deltatheta_peaks}
\end{figure}

The dynamical equations for the variables $\delta$ and $\Theta$ are given by 
\begin{subequations}
\label{ell_phasespace}
\begin{align} 
\frac{d \delta_{\mathrm{ell}}}{d \ln a} &=-(1+ \delta_{\mathrm{ell}}) \Theta_{\mathrm{ell}}\\
\label{veldivell}\frac{d \Theta_{\mathrm{ell}}}{d \ln a} &= -\frac{1}{2}\left[3 \Omega_m(a) \delta_{\mathrm{ell}}  - \left\{ \Omega_m(a) - 2 \Omega_\Lambda(a)-2 \right \}  \Theta_{\mathrm{ell}}
+ \frac{2}{3} \Theta_{\mathrm{ell}}^2 + 2 \sigma^2 \right], 
\end{align}
\end{subequations}
where 
\beq
\nonumber \sigma^2 = \sum_i \sigma_i^2;  \; \; \; \sigma_i = \lambda_{v,i} - \Theta_{\mathrm{ell}}/3.
\label{shearscalar}
\eeq 
$\sigma_i$s are the eigenvalues of the traceless `shear' component of the tensor of velocity derivatives.
The $\delta$ equation follows from the definitions in \eqnrefs{deltadef}, \eqnrefbare{lvdef} and \eqnrefbare{Thetadef}. The $\Theta$ equation follows from summing \eqnref{dyn2} over all $i$ and using the relation $\sum_i \lambda_{v,i}^2 = \sigma^2 + \Theta_{\mathrm{ell}}^2/3$. The corresponding equations for the sphere are (N13)
\begin{subequations}
\label{sphere_phasespace}
\begin{align} 
\label{ddeltasph}\frac{d \delta_{\mathrm{sph}}}{d \ln a} &= -(1+ \delta_{\mathrm{sph}}) \Theta_{\mathrm{sph}}\\
\label{dthetasph}\frac{d \Theta_{\mathrm{sph}}}{d \ln a} &= -\frac{1}{2}\left[  3 \Omega_m(a)  \delta_{\mathrm{sph}} -  \left\{ \Omega_m(a) - 2 \Omega_\Lambda(a)-2 \right \} \Theta_{\mathrm{sph}} + \frac{2}{3} \Theta_{\mathrm{sph}}^2 \right]. 
\end{align}
\end{subequations}
\capfigref{deltatheta} shows $\{\delta, \Theta\}$ pairs from one realization of the case $\sigma_G=1$. The dashed line is the prediction of linear theory given in \eqnref{dvdrlin}. The solid black line is the SC-DVDR (eq. \ref{eqsphere}) and it acts as limiting case for the dynamics. We can prove this mathematically by the following arguments. 
Suppose that at some instant of time the perturbation pairs of the ellipsoid and sphere are the same i.e., $\{\delta_{\mathrm{ell}},\Theta_{\mathrm{ell}}\} = \{\delta_{\mathrm{sph}}, \Theta_{\mathrm{sph}}\}$ and hence both lie on the Zeldovich curve of the sphere. The ellipsoidal pair is governed by \eqnref{ell_phasespace} where as the spherical pair is governed by \eqnref{sphere_phasespace}. The rate of evolution differs in the two systems; the difference is 
\beq
\left.\frac{d \Theta}{d\delta}\right|_{\mathrm{ell}} - \left.\frac{d \Theta}{d\delta}\right|_{\mathrm{sph}} =   \frac{\sigma^2}{(1+ \delta) \Theta}.
\label{differences}
\eeq
The terms $(1+\delta)$ and $\sigma^2$ are both positive.  
Thus, when $\Theta<0$, i.e., the region is contracting, $\left.\frac{d\Theta}{d\delta}\right|_{\mathrm{ell}} < \left.\frac{d\Theta}{d\delta}\right|_{\mathrm{sph}}$ and when $\Theta$ is positive 
 $\left.\frac{d\Theta}{d\delta}\right|_{\mathrm{ell}} > \left.\frac{d\Theta}{d\delta}\right|_{\mathrm{sph}}$. Since the slope of the SC-DVDR i.e., $\frac{d\Theta}{d\delta}_{\mathrm{sph}}$ is always negative, these inequalities imply that  triaxiality causes overdense regions to infall faster and underdense regions to expand slower. Triaxiality also implies some regions which are underdense but infalling, a possibility excluded in the spherical model. 

It is interesting to note that the linear relation is a good approximation for the mean behaviour at late times. This can be partly explained by comparing with \figref{lambdadandv}. The mean relation for the largest (in magnitude) eigenvalues is also close to linear. In this limit the collapse can be modelled as a 1D infall along the largest eigendirection, for which the density-velocity relation is linear (Appendix \ref{1ddynamics}).

Numerical simulations solve for the dynamics almost exactly, accounting for a variety of other effects such as non-local physics due to interaction between the environment, accretion etc. However, it is found that the spherical collapse prediction fits the `mean' $\delta-\Theta$ relation reasonably well \citep{bernardeau_non-linearity_1999, kudlicki_reconstructing_2000,bilicki_velocity-density_2008}. On the other hand, predictions from the ellipsoidal collapse model systematically differ from this relation because of the shear term (see eq. \ref{ell_phasespace}). This seemingly better agreement of the spherical collapse model can be roughly explained as follows. The general evolution of the velocity divergence, given by Raychaudhuri's equation, depends on the shear $\sigma^2$ and vorticity ($\frac{1}{2} {\bmath \omega} \cdot {\bmath \omega}$). These two terms arise with opposite signs (see \citealt{peebles80}, eq. 22.14); for overdensities, shear speeds up evolution of $\Theta$ towards negative values and vorticity does the opposite, possibly causing a partial cancellation at late times. The non-rotating ellipsoidal model includes shear but not vorticity, whereas the spherical collapse includes neither, mimicking the partial cancellation. This argument, while plausible, requires further investigations, but these are beyond the scope of the current paper. The predictions of ellipsoidal collapse agree better if one considers only the peaks (since haloes in simulations form at peaks in the density field). For the case of homogenous triaxial collapse, this corresponds to initial conditions where all the three $\lambda_d$s are positive (an infalling halo) or all three are negative (an ever expanding void). \capfigref{deltatheta_peaks} shows the phase space snapshot for such special initial conditions evolved using  \eqnrefs{phspdyn}. Clearly, the scatter decreases and the points lie closer to the spherical or `mean' relation.

\subsection{Scatter in the $\delta-\Theta$ plane}
\label{sec:scatter}
\begin{figure}
\includegraphics[width=18cm]{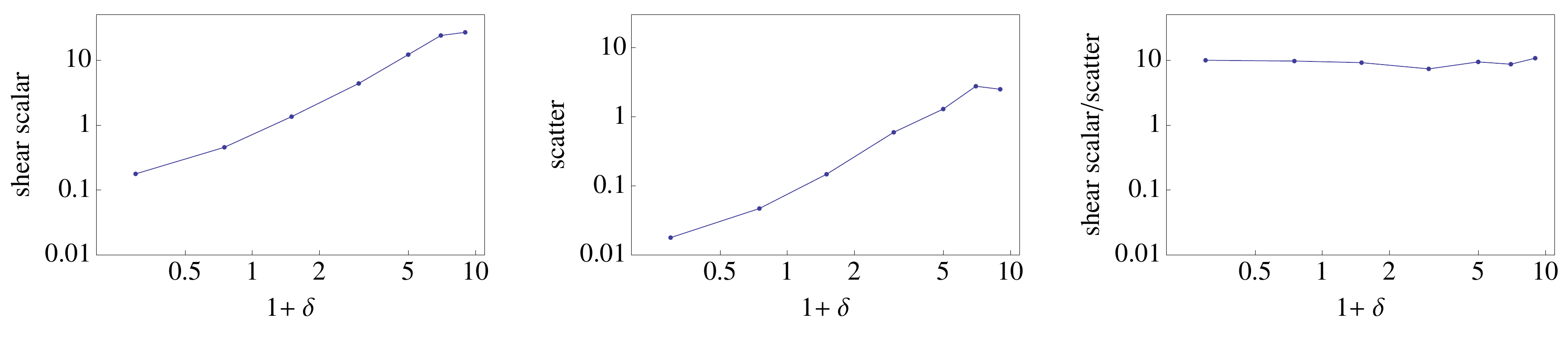}
\caption{Shear and scatter: the left, middle and right panels show the conditional shear scalar, scatter and their ratio as a function of the density. The conditional shear scalar defined as $\left \langle\sigma^2\right\rangle_{|\delta}$ and the conditional scatter as $\left \langle \left (\Theta - \langle \Theta\rangle_{|\delta}\right)^2 \right \rangle _{|\delta}$. Both are minimal in the void limit and increase as the density increases. Their ratio is roughly constant indicating that the scatter is related to the non-zero shear, which arises due to the asymmetry in the system. }  
\label{scatter}
\end{figure}
The spread in the $\delta-\Theta$ relation is least at the void limit $\delta=-1$ and increases as $\delta \rightarrow \infty$. This means that the void limit of the ellipse is the same as the sphere and we can compute it analytically. 
From \eqnref{eqsphere} this is 
\beq 
\Theta_{\mathrm{max}} = \left.\Theta\right|_{\delta = -1} = \frac{3}{2} \Omega_m^{0.56}.
\label{thetamax}
\eeq
We note that the exponent of $\Omega_m$ is slightly different from the value of $0.6$ given in previous works (\citealt{bernardeau_omega_1997, bernardeau_quasi-gaussian_1992}). This is important for proposals to constrain the value of $\Omega_m$ using the value of maximal expansion in voids (for e.g. \citealt{dekel_omega_1994}).

The non-linear density-velocity divergence relation (DVDR), also known as the gravity-velocity relation, has been discussed  in great detail by various authors in the past (\citealt{bernardeau_quasi-gaussian_1992, nusser_cosmological_1991,gramann_second-order_1993, bernardeau_new_1996, chodorowski_weakly_1997, chodorowski_recovery_1998, bernardeau_non-linearity_1999,kudlicki_reconstructing_2000,ciecielg_gaussianity_2003,kitaura_estimating_2012}). These analyses are based on higher order perturbation theory (see review \citep{bernardeau_large-scale_2002}), either in Eulerian or Lagrangian frame, or on numerical simulations. All of them show a scatter around the mean-relation which is usually attributed to two reasons. One reason is the non-locality of the dynamics captured by higher order perturbation theory. In linear theory, both in Eulerian space (continuity equation) or in Lagrangian space (Zeldovich approximation), the DVDR relation is local i.e., the velocity divergence/velocity at a point is given by the density/gravitational acceleration at that point. Higher orders of perturbation theory depend on derivatives of lower orders and to determine them, the field needs to be known everywhere (non-local). This non-locality makes the relation stochastic or non-deterministic because although the field equations are deterministic, the initial conditions are random. Another related reason is the traceless `shear' component of the tensor of velocity derivatives. The eigenvalues of this tensor are $\sigma_i = \lambda_{v,i} - \Theta/3$ and their relation to the scatter has been discussed by \cite{chodorowski_large-scale_1997} from the point of view of perturbation theory. The triaxial model provides a nice illustration of the same effect based on local dynamics. This is clear from considering \eqnref{differences}. The terms that differ from the sphere are all second order and vanish when all the axes are equal. \capfigref{scatter} illustrates this effect for the data plotted in \figref{deltatheta}. The left panel plots the (conditional) shear scalar, $\left \langle\sigma^2\right\rangle_{|\delta}$, where $\sigma^2$ is defined in \eqnref{shearscalar}. The shear is small in the void regions and increases in the overdense regions. The middle panel shows the conditional scatter defined as $\left \langle \left(\Theta - \langle \Theta  \rangle_{|\delta}\right)^2 \right \rangle _{|\delta}$. This has the same trend as that of the conditional shear.  The last panel shows their ratio. Interestingly, this ratio is roughly constant over two decades in density \footnote{The ratio is sensitive to the binning in high density regions, but is of the same order of magnitude.}. Thus, the scatter can be attributed to the asymmetry in the system. Indeed, the non-linear DVDR based on the spherical collapse model, which is both local and symmetric, and shows no scatter. 


\section{Marginal Probabilities $\MakeLowercase{p}(\Theta)$ and $\MakeLowercase{p}(\delta)$. }
\label{sec:application}
As an application of this method, we compute the marginal probabilities $p(\Theta)$ and $p(\delta)$ and qualitatively discuss the joint PDF $p(\delta, \Theta)$. 
\subsection{Numerical runs}
\label{numrun}
The numerical runs were performed by evolving a set of $10^4$ initial conditions using the equations of phase space dynamics given in \eqnref{phspdyn}. At the desired final time the $\delta$ and $\Theta$ are computed according to definitions given in \eqnrefs{delta_ld} and \eqnrefbare{Thetadef}. Each set was drawn from the distribution given by \eqnref{pdf}. Three scales were considered: $\sigma_G=0.5,1$ and 2. The initial $\sigma_G$ is related to the scale of the perturbation: the exact dependence depends on the shape and amplitude of the power-spectrum. For the BBKS power spectrum with $n_s=1$ and $\sigma_8=0.9$, $\sigma_G=0.5,1,2$ corresponds to length scales of $R_f =$ 16.4,7 and 3.65 $h^{-1}$Mpc respectively. Two cosmologies were considered: EdS ($\Omega_m=1, \Omega_\Lambda=0$) and $\Lambda{\rm CDM}$ $(\Omega_m=0.29, \Omega_\Lambda=0.71$). The realization at $a=1$ was the same for the two cosmologies, but the values at the initial epoch $a=0.001$ were set by multiplying by the correct growth rate factor. By $a=1$ roughly one-tenth of the points had undergone collapse.  For each $\sigma_G$ value, five realizations were evolved; the PDF is the average over five and the error bars correspond to the standard deviation. 
 
\subsection{Theoretical Estimates for comparison}
The one-point distributions of the non-linear density and velocity fields have been discussed in great detail in the past. One of the most popular forms for the density PDF is the empirically motivated log-normal model given by \cite{coles_lognormal_1991}. While this form has been checked by simulations (for e.g., \citealt*{kayo_probability_2001}, \citealt{kitaura_estimating_2012}), there are others based on more analytical approaches. For example, \cite{kofman_evolution_1994} constructed the non-linear PDF from the linear Gaussian PDF by expressing the non-linear density as a function of the linear $\lambda_d$ via the Zeldovich approximation. Further work by Bernardeau \& collaborators (\citealt{bernardeau_effects_1994, kofman_evolution_1994}) gave forms for the large scale density PDF, both in Eulerian and Lagrangian spaces, from cumulants calculated using perturbation theory. Later \cite{fosalba_cosmological_1998-1} gave an alternative approach to computing the cumulants using the spherical collapse model as a local approximation for the dynamics. 
\cite*{ohta_evolution_2003,ohta_cosmological_2004} formulated the differential equations for the evolution of the one-point PDFs and solved them using the spherical and ellipsoidal collapse as local approximations for the dynamics. More recently, \cite*{lam_ellipsoidal_2008,lam_perturbation_2008} derived the density PDF both in real space and redshift space based on excursion sets and ellipsoidal collapse. 

For the density comparison we choose a combination of the log-normal forms and the perturbative form proposed by 
\cite{bernardeau_effects_1994}:
\beq
\label{theopdf}
p(\delta) = \left \{\begin{array}{cc}
p_{B94}^{void}(\delta) &-1\leq \delta < -0.4 \\\\
p_{L-N}(\delta) &-0.4\leq\delta < 1 \\\\
 p_{B94}^{high}(\delta) &\delta\geq1,
\end{array}
\right.
\eeq
where
\bea
\label{B94void}
p_{B94}^{void}(\delta) d\delta&=&\left(\frac{7-5(1+ \delta)^{2/3}}{4 \pi \sigma_\delta^2}\right)^{1/2} (1+\delta)^{-5/3}  \times \exp\left[-\frac{9}{8\sigma_\delta^2}\left(-1+\frac{1}{(1+\delta)^{2/3}}\right)^2 \right] d\delta\\ \nonumber \\
\label{lognormal} p_{L-N}(\delta) d\delta &=& \frac{1}{\sqrt{2 \pi} (1+\delta) \sigma_{ln}} \times\exp\left[- \frac{\{\log(1+\delta) + \sigma_{ln}^2/2\}^2}{2 \sigma_{ln}^2}\right] d \delta \\ \nonumber \\
p_{B94}^{high}(\delta) d\delta &=& f_c \frac{3 a_{s\delta} \sigma_\delta}{4 \sqrt{\pi}} (1+\delta)^{-5/2} \times \exp\left[\frac{-|y_{s\delta}| \delta+ |\phi_{s\delta}|}{\sigma_\delta^2 }\right] d\delta 
\label{b94high}
\eea
with $\sigma_{ln} = \ln(1+\sigma_\delta^2)$, $a_{s\delta}=1.84, y_{s\delta} =-0.184, \phi_{s\delta} = -0.03$. We have chosen the $n=-3$ values for the parameters $a_s,y_s, \phi_s$ from B94 (this corresponds to the case of no smoothing). The correction factor $f_c =  [1 +2 (0.8 -\sigma_\delta)\sigma_\delta^{-1.3} (1+\delta)^{-0.5}]$ was introduced by B94 to account for the fact that the PDF calculated by the perturbative form did not perform well at high $\delta$.

For the velocity divergence the form for $p(\Theta)$ proposed in B94 
\beq
\label{theopdfv}
p(\Theta) = \left \{\begin{array}{cc}
p_{B94}^{void}(\Theta) & 1.5 \geq \Theta \geq -0.5 \\ \\
p_{B94}(\Theta) & \Theta < -0.5 
\end{array}
\right.
\eeq
where
\bea 
p_{B94}^{void}(\Theta)d \Theta  &=& \frac{1}{3} (3+2\tau)^2 \sqrt{\frac{1+2\tau}{3 \pi \sigma_\Theta^2 (3+2  \tau)}}  \times \exp\left(- \frac{\tau^2}{2 \sigma_\Theta^2}\right) d \Theta; \; \; \;\;  \tau = \Theta \left(1 - \frac{3}{2} \Theta\right )^{-1}\\\nonumber \\
p_{B94}^{high}(\Theta) d\Theta &=& f_c \frac{3 a_{s\Theta} \sigma_\Theta}{4 \sqrt{\pi}} \left(\frac{3}{2} - \Theta\right)^{-5/2} \times \exp\left[\frac{|y_{s\Theta}| \Theta+ |\phi_{s\Theta}|}{\sigma_\delta^2 }\right] d\Theta
\label{B94hightheta}
\eea
with $a_{s\Theta}=1.67, y_{s\Theta} = -0.222, \phi_{s\Theta} = -0.042$. 
The PDF for the $\Lambda$CDM cosmology is given by rescaling $p_{\Lambda CDM}(\Theta)  = p_{EdS}(\Theta \rightarrow \Theta/f(\Omega_m))$, where the $\sigma_\Theta$ is the variance of the scaled variable. The correction factor $f_c = [1 + 30 (0.8 -\sigma_\Theta)\sigma_\Theta^{-1.3} (1.5-\Theta)^{-0.5}]$ was introduced in B94 to account for the fact that the expression \eqnref{B94hightheta} underestimates the exact answer. We used it only for the case $\sigma_G=1, a=1$.  For all other cases, $f_c=1$. 

The $\sigma_\delta$ and $\sigma_\Theta$ in the above expressions are related to $\sigma_G$ using linear theory $\sigma_\delta^2(a) = \sigma_G^2 D_+^2(a)/D_+^2(a=1)$ and not from data. Ideally, these expressions are valid for small values of $\sigma$. So we do not compare the case $\sigma_G=2$ at $a=1$. 

 The PDFs generated by our analysis are in the Lagrangian frame since the evolution of the ellipse conserves mass and the density is related to the change in volume. For a fair comparison to the forms discussed above one must convert from the Lagrangian frame to the Eulerian frame. We follow a procedure along the lines discussed in \cite{bernardeau_effects_1994}; however, the correct method involves taking derivatives of the Lagrangian PDFs with respect to the mass scale (i.e., varying $\sigma_G$). Since we have used only three values of $\sigma$, this computation will be highly inaccurate. Instead, we use the linear limit of the relation which, for the case $n=-3$, reads (see appendix \ref{EtoLframe})
\bea
p_E (\delta) &=& \frac{p_L(\delta)}{1+\delta} \\
p_E(\Theta) &=& \frac{p_L(\Theta)}{1+\delta}, 
 \eea
 where $p_E$ and $p_L$ are the Eulerian and Lagrangian PDFs respectively. This correction is applied for each realization and the average is computed over five realizations. The PDFs in either frame are not a priori normalized \footnote{Numerically we found that they were normalized with an error of a few percent}. We numerically normalize the Eulerian PDF over the range of values considered.

\subsection{Results: $p(\delta)$ and $p(\Theta)$}
We construct PDFs from the data generated by the numerical runs described in \S \ref{numrun}. The output is analysed at three different epochs: $a=0.05,0.4\;  {\rm and}\; 1$ (corresponding to $z=19,1.5\; {\rm and }\; 0$). Three scales were considered: $\sigma_G=0.5$ (red),1 (blue dotted), 2 (brown, dashed).  

\begin{figure}
\includegraphics[width=17cm]{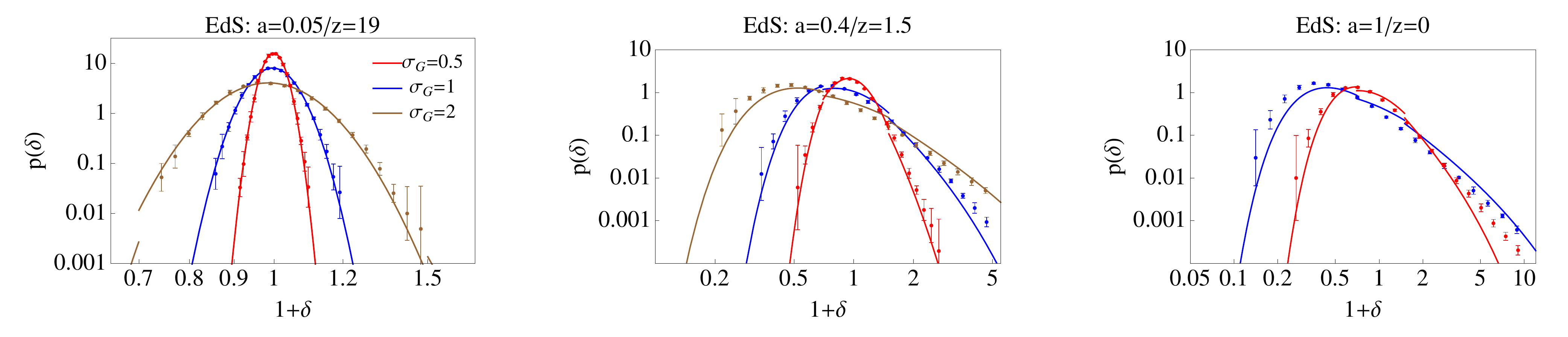}\\
\includegraphics[width=17cm]{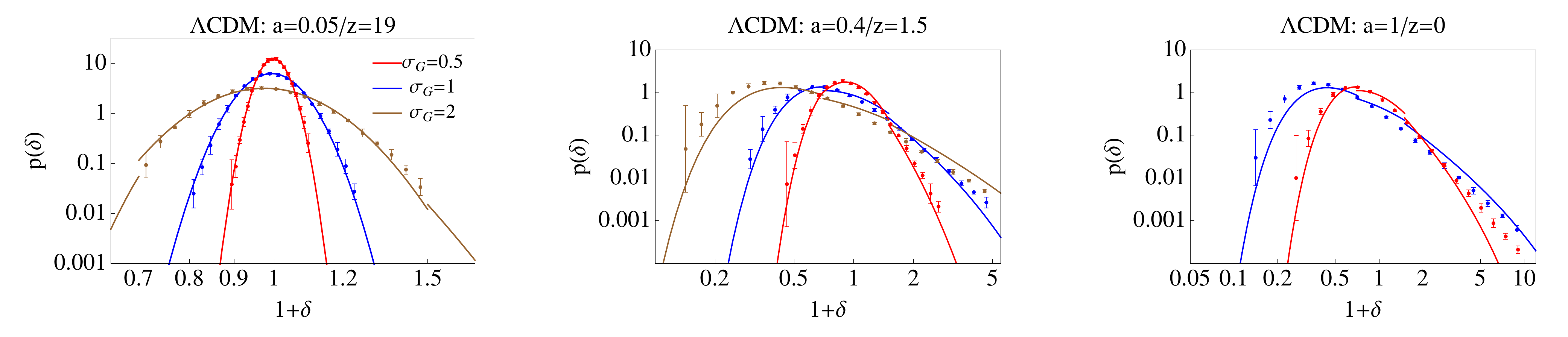}
\caption{Evolution of $p(\delta)$ vs $1+\delta$ for three values of $\sigma_G$ for the EdS and $\Lambda$CDM models. Agreement with the theoretical forms given in \eqnref{theopdf} is better in the void regions than in the high density tails.  }
\label{marginalden}
\end{figure}
\capfigref{marginalden} shows the marginal probability distribution $p(\delta)$ at the three epochs. The points denote the data and the lines denotes the analytic distribution. The top and bottom panels are the EdS and $\Lambda$CDM cosmologies. At early times, the departure from Gaussianity is small and the log-normal model provides a good fit. At later epochs, the fits are better for smaller $\sigma_G$. This departure can be attributed to several reasons. Firstly, triaxial collapse is local, whereas higher order perturbation theory can account for non-local effects. The other reason is be that the perturbative approximation fails at high values of $\delta$ and the correction factors given in B94 are expected to work only for $\sigma \lesssim 1$. Thirdly, we are using a linear approximation for the transformation from the Lagrangian to Eulerian frame. 

\begin{figure} 
\includegraphics[width=17cm]{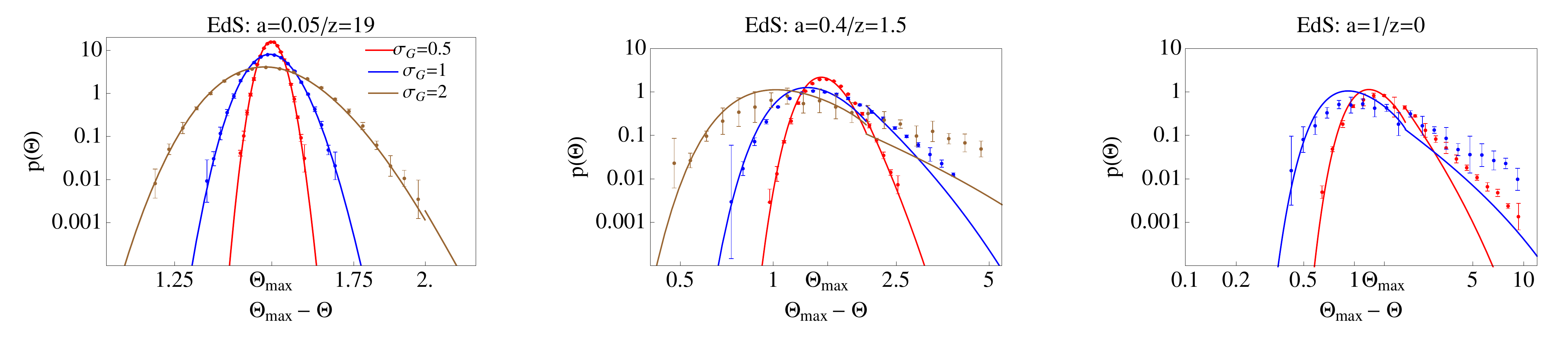}\\
\includegraphics[width=17cm]{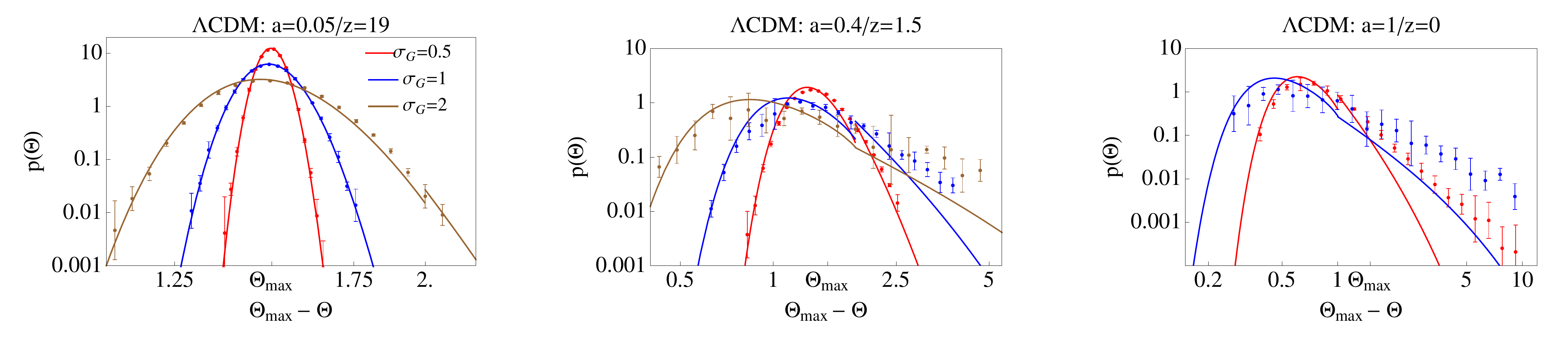}
\caption{Evolution of $p(\Theta)$ vs $\Theta_{\mathrm{max}} - \Theta$, where $\Theta_{\mathrm{max}} = 1.5 \Omega_m^{0.56}$. For EdS, $\Theta_{\mathrm{max}}$ is constant with time, where as for $\Lambda$CDM it changes. Agreement with the theoretical forms of B94 given in \eqnref{theopdfv} is maintained in voids throughout the evolution, but worsens in the overdense regions $\Theta_{\mathrm{max}} - \Theta \gg1$. }
\label{marginalvel}
\end{figure}

The PDF of $\Theta$, shown in \figref{marginalvel}, exhibits a similar behaviour. The $x$-axis in this case is $\Theta_{\mathrm{max}}-\Theta$, where $\Theta_{\mathrm{max}}$ is the maximal value in voids given by \eqnref{thetamax}. The point $\Theta_{\mathrm{max}}$ in this variable corresponds to $\Theta=0$ and is shown on the graph. Points to the left of $\Theta_{\mathrm{max}}$ correspond to voids and those to the right correspond to overdensities. We see that the analytic expressions given by  \eqnref{theopdfv} fit very well at early times when the perturbations are in the linear regime and the agreement reduces with increasing epoch and $\sigma$. 

\subsection{Joint pdf} 
\begin{figure}
\includegraphics[width=16cm]{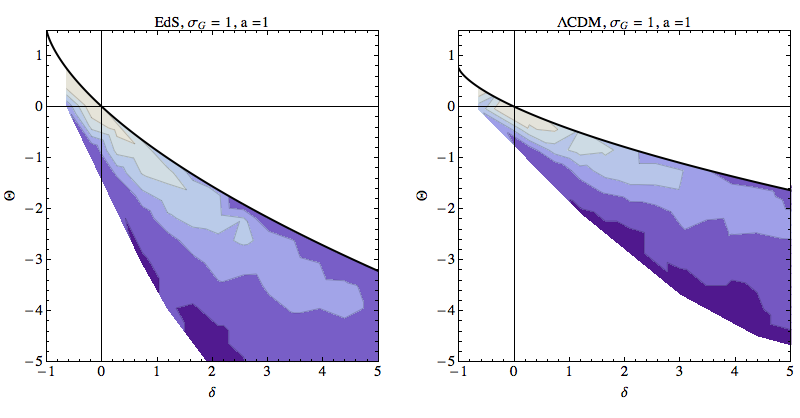}
\caption{The joint PDF $p(\delta, \Theta)$. The contours are drawn for $p(\delta, \Theta) =0.32,0.1,0.032,0.01,0.001$. The SC-DVDR, given by \eqnref{eqsphere}, forms a limiting case for the joint PDF. The spread is related to the shear component of the velocity field as was discussed in \S \ref{sec:scatter} }
\label{jointpdf}
\end{figure}

\capfigref{jointpdf} shows the evolution of the joint PDF $p(\delta, \Theta)$ for the EdS model (left panel) and the $\Lambda$CDM model (right panel) at $a=1$. The region above the SC-DVDR is overlaid to indicate the sharp cut-off observed in \figref{deltatheta}. Our graphs are in reasonably good agreement with the joint PDF plotted by \cite*{ohta_evolution_2003}, which was also based on ellipsoidal collapse, although they have a slight spill over which may possibly be a plotting artefact. Both for EdS and $\Lambda$CDM, the scatter increases as time elapses (at early times the scatter is minimal; all points obey the local linear relation). However, the limiting spherical curve in the two cases is different. The scatter is greater for a higher $\sigma_G$ (not shown).  In N13, it was demonstrated how the attracting nature of the SC-DVDR could be exploited to remove parameter degeneracies related to the power spectrum normalization or index. But the same cannot be done with the joint PDF. Unlike the SC-DVDR which does not depend on the initial $\sigma_G$, the joint PDF carries the signature of the initial width (we investigated two more values of $\sigma_G$; data not shown). 

\section{Conclusion}
 \label{sec:conclusion}
 
The main points of this paper can be summarized as follows. 
\begin{itemize}

\item We recast the Bond \& Myers' set of equations for triaxial collapse into a new set of equations governing the dynamics of eigenvalues of the deformation tensor ($\lambda_{a,i}$), the velocity derivative tensor ($\lambda_{v,i}$) and the gravity Hessian ($\lambda_{d,i}$). The main advantage of this reform is that it eliminates the dependence on complicated elliptic integrals which are present in the original system and provides a more natural way to track the dynamics of the perturbation fields. With these definitions $\delta = \sum \lambda_{d,i}$ and the scaled velocity divergence $\Theta = \sum \lambda_{v,i}$, $i= 1,2,3$. 

\item In linear theory `no decaying modes' implies that the $\lambda_d$ and $\lambda_v$ for each axis are proportional, but this breaks down in the non-linear regime. Using the ideas in N13, we extended this to the non-linear regime by imposing the condition `no perturbations at the big bang'. This gives a relation between the three $\lambda_d$ and $\lambda_v$s at any epoch  which traces out a 3D subspace of the 6D perturbation space. We find that along this subspace the perturbations satisfy 
\beq 
{\tilde q}_v + {\tilde q}_d=1,
 \eeq
where ${\tilde q} = (\lambda_{\mathrm{max}} - \lambda_{\mathrm{inter}})/(\lambda_{\mathrm{max}} - \lambda_{\mathrm{min}}).$
To the best of our knowledge, this is a new universal relation, not discussed in earlier literature. 
Since the dependence on cosmology of the velocity-gravity relation is primarily through the linear growth factor, this universality is with respect to redshift as well as cosmological model. 

\item Using the same phase space equations, we analysed the relation between the density $\delta$ and the trace of the deformation tensor $\sum \lambda_{a,i}$. In the linear regime, $\lambda_{d,i} \approx \lambda_{a,i}$ and $\delta \approx \sum \lambda_{a,i}$. 
To understand the validity of this statement we examined the $\delta-\sum \lambda_{a,i}$ relation a function of time. We find that linearity of this relation breaks down at late times even for `linear' $\delta$ values. This emphasizes the point that the `linearity' refers to the individual $\lambda$s. In the non-linear regime, because of a cancellation between $\lambda_a$s of opposite signs, it is possible to have a `linear' $\delta$ although the $\lambda$s are of order unity. 

\item From the nine-dimensional set for the eigenvalues, we obtained a two dimensional set that governs the $\delta-\Theta$ dynamics. We find that the late time density-velocity divergence relation is close to linear. In this regime, the collapse along the shortest axis dominates the collapse along the other two ones. Therefore, the dynamics can be effectively modelled as a one dimensional with the other two axes comoving with the background.

\item When compared to numerical simulations (for e.g., \citealt{bernardeau_non-linearity_1999,kudlicki_reconstructing_2000}), triaxial collapse does worse than the spherical collapse
in predicting the $\delta-\Theta$ relation. Both triaxial collapse and simulations show a scatter in the relation, but in the first case the spherical relation acts as a upper bound whereas in the case of simulations it turns out to be a good approximation for the mean. In the case of simulations, the scatter arises from random errors coupled with stochastic initial conditions and it is somewhat of a coincidence that the mean relation is well described by the sphere. On the other hand, the scatter in the ellipsoidal model, arises due to the asymmetry in the system and is related to the `shear' part of the velocity derivative tensor. This effect has been discussed in the past by \cite{chodorowski_large-scale_1997} using higher order perturbation theory and the triaxial collapse provides a nice illustration of the same based on `local' dynamics. 

\item As an application of this method, we examined the pdf of the density and velocity perturbations. We focussed primarily on the marginal PDFs $p(\Theta)$ and $p(\delta)$ and discussed the joint PDF only qualitatively. We find that the agreement is good at earlier epochs and smaller values of $\sigma_G$. This is somewhat expected. The theoretical estimates based on perturbation theory include non-local physics whereas the ellipsoidal model is local. In addition, the conversion from the Lagrangian frame to Eulerian frame is based on a linear relation, which will breakdown for high values of $\delta$. As another application we will consider the evolution of axis ratios (paper II, \citealt*{nadkarni-ghosh_phase_2015}). 

\end{itemize}

The aim of this work was to understand the non-linear behaviour of density and velocity perturbations through the eigenvalue dynamics. This work is general and gives rise to a range of possible applications. In particular, there is recent interest in numerically classifying and quantifying the cosmic web based on eigenvalues of the gravity Hessian ({\citealt{hahn_properties_2007, forero-romero_dynamical_2009}) and the eigenvalues of the velocity derivative tensor (\citealt{hoffman_kinematic_2012}; \citealt*{libeskind_velocity_2014}).One main issue in these studies is that the resultant mathematical structure depends not only on whether one uses the gravitational tensor (T-web) or the velocity tensor (V-web), but it also depends on the details of the classification algorithm (\citealt{hoffman_kinematic_2012}; see \citealt{forero-romero_cosmic_2014} for a list). Furthermore, it was also found that the velocity based classification is a better tracer of the cosmic web and there have been independent observational evidences for the same (\citealt*{lee_alignments_2014}). However, the detailed reasoning based on a first principles analysis for this connection is unclear. Through the newly defined `growth enhancement factor' we are able to characterize the growth rates and provide an insight into this result. Further investigations based on eigenvalue dynamics may help understand these numerical studies better. 
Another possible application is to improve ellipsoidal collapse based mass function generating codes, such as PINOCCHIO (\citealt{monaco_accurate_2013}). This will be useful to investigate the universal nature of mass function, dependence on cosmology etc. 
Analytic descriptions of triaxial dynamics can also be used to resolve issues related to alignment or initial shapes of haloes which have been raised in recent simulations (see for e.g. \citealt*{despali_ellipsoidal_2013}) because time-reversing the phase flow equations is easy making it possible to trace back to the initial conditions. 

However, there are many limitations of the triaxial model considered here. The first important assumption is that the principle axes stay fixed throughout the evolution i.e., no rotations are included. Second, the ellipse is isolated; the dynamics is local. There are no interactions between neighbours. The effect of the environment is modelled through the effective external non-linear tidal tensor, which depends completely on the axes lengths. The first extension of this analysis would be to follow the more complete set of equations such as those given by \cite{eisenstein_analytical_1995}, which includes rotation. Additional parameters will be required to model the rotational degree of freedom, but the basic framework remains the same. The long-term aim of such investigations would be to provide analytic insight that helps to interpret and improve numerical studies. This paper presents a first step towards this goal. 

\section{Acknowledgements}
We would like to thank the referee, Micha{\l} Chodorowski, for his detailed comments and suggestions which improved the original manuscript significantly. In addition, we would like to thank Sagar Chakraborty for useful discussions regarding dynamical systems. S.N. would like to thank the hospitality of ICTP, Trieste, and discussions with Pierluigi Monaco and Ravi Sheth which sowed the seeds of this work. Thanks are also due to Pierluigi Monaco for a careful reading of the manuscript and valuable suggestions. 

\clearpage
\bibliographystyle{mn2e.bst}
\bibliography{ellbibtex1,ellbibtex2,ellbibtex3}

\begin{thebibliography}{83}
\expandafter\ifx\csname natexlab\endcsname\relax\def\natexlab#1{#1}\fi

\bibitem[{Angrick(2013)}]{angrick_ellipticity_2013}
Angrick C., 2013, {ArXiv} e-prints, 1305, 497

\bibitem[{Angrick \& Bartelmann(2010)}]{angrick_triaxial_2010}
Angrick C., Bartelmann M., 2010, Astronomy and Astrophysics, 518, 38

\bibitem[{Audit {et~al}\mbox{.}(1997)Audit, Teyssier, \&
  Alimi}]{audit_non-linear_1997}
Audit E., Teyssier R., Alimi J.-M., 1997, Astronomy and Astrophysics, 325, 439

\bibitem[{Bagla \& Padmanabhan(1996)}]{bagla_new_1996}
Bagla J.~S., Padmanabhan T., 1996, The Astrophysical Journal, 469, 470

\bibitem[{Bardeen {et~al}\mbox{.}(1986)Bardeen, Bond, Kaiser, \&
  Szalay}]{bardeen_statistics_1986}
Bardeen J.~M., Bond J.~R., Kaiser N., Szalay A.~S., 1986, The Astrophysical
  Journal, 304, 15

\bibitem[{Bernardeau(1992)}]{bernardeau_quasi-gaussian_1992}
Bernardeau F., 1992, The Astrophysical Journal, 390, L61

\bibitem[{Bernardeau(1994)}]{bernardeau_effects_1994}
Bernardeau F., 1994, Astronomy and Astrophysics, 291, 697

\bibitem[{Bernardeau {et~al}\mbox{.}(1999)Bernardeau, Chodorowski, {\L}okas,
  Stompor, \& Kudlicki}]{bernardeau_non-linearity_1999}
Bernardeau F., Chodorowski M.~J., {\L}okas E.~L., Stompor R., Kudlicki A.,
  1999, Monthly Notices of the Royal Astronomical Society, 309, 543

\bibitem[{Bernardeau {et~al}\mbox{.}(2002)Bernardeau, Colombi, Gazta�aga, \&
  Scoccimarro}]{bernardeau_large-scale_2002}
Bernardeau F., Colombi S., Gazta�aga E., Scoccimarro R., 2002, Physics
  Reports, 367, 1

\bibitem[{Bernardeau \& van~de Weygaert(1996)}]{bernardeau_new_1996}
Bernardeau F., van~de Weygaert R., 1996, Monthly Notices of the Royal
  Astronomical Society, 279, 693

\bibitem[{Bernardeau {et~al}\mbox{.}(1997)Bernardeau, van~de Weygaert, Hivon,
  \& Bouchet}]{bernardeau_omega_1997}
Bernardeau F., van~de Weygaert R., Hivon E., Bouchet F.~R., 1997, Monthly
  Notices of the Royal Astronomical Society, 290, 566

\bibitem[{Bilicki \& Chodorowski(2008)}]{bilicki_velocity-density_2008}
Bilicki M., Chodorowski M.~J., 2008, Monthly Notices of the Royal Astronomical
  Society, 391, 1796

\bibitem[{Bonamigo {et~al}\mbox{.}(2014)Bonamigo, Despali, Limousin, Angulo,
  Giocoli, \& Soucail}]{bonamigo_universality_2014}
Bonamigo M., Despali G., Limousin M., Angulo R., Giocoli C., Soucail G., 2014,
  {arXiv}:1410.0015 [astro-ph], {arXiv}: 1410.0015

\bibitem[{Bond \& Myers(1996)}]{bond_peak-patch_1996}
Bond J.~R., Myers S.~T., 1996, Astrophysical Journal Supplement Series, 103, 1

\bibitem[{Buchert(1992)}]{buchert_lagrangian_1992}
Buchert T., 1992, Monthly Notices of the Royal Astronomical Society, 254, 729

\bibitem[{Carlson(1987)}]{carlson87}
Carlson B., 1987, Mathematics of Computation, 49, 595

\bibitem[{Carlson(1989)}]{carlson89}
Carlson B., 1989, Mathematics of Computation, 53, 327

\bibitem[{Chodorowski(1997)}]{chodorowski_large-scale_1997}
Chodorowski M.~J., 1997, Monthly Notices of the Royal Astronomical Society,
  292, 695

\bibitem[{Chodorowski \& {\L}okas(1997)}]{chodorowski_weakly_1997}
Chodorowski M.~J., {\L}okas E.~L., 1997, Monthly Notices of the Royal
  Astronomical Society, 287, 591

\bibitem[{Chodorowski {et~al}\mbox{.}(1998)Chodorowski, {\L}okas, Pollo, \&
  Nusser}]{chodorowski_recovery_1998}
Chodorowski M.~J., {\L}okas E.~L., Pollo A., Nusser A., 1998, Monthly Notices
  of the Royal Astronomical Society, 300, 1027

\bibitem[{Ciecielag {et~al}\mbox{.}(2003)Ciecielag, Chodorowski, Kiraga,
  Strauss, Kudlicki, \& Bouchet}]{ciecielg_gaussianity_2003}
Ciecielag P., Chodorowski M.~J., Kiraga M., Strauss M.~A., Kudlicki A., Bouchet
  F.~R., 2003, Monthly Notices of the Royal Astronomical Society, 339, 641

\bibitem[{Coles \& Jones(1991)}]{coles_lognormal_1991}
Coles P., Jones B., 1991, Monthly Notices of the Royal Astronomical Society,
  248, 1

\bibitem[{Dekel \& Rees(1994)}]{dekel_omega_1994}
Dekel A., Rees M.~J., 1994, The Astrophysical Journal Letters, 422, L1

\bibitem[{Del~Popolo {et~al}\mbox{.}(2001)Del~Popolo, Ercan, \&
  Xia}]{del_popolo_ellipsoidal_2001}
Del~Popolo A., Ercan E.~N., Xia Z., 2001, The Astronomical Journal, 122, 487

\bibitem[{Del~Popolo \& Gambera(2000)}]{del_popolo_non_2000}
Del~Popolo A., Gambera M., 2000, Astronomy and Astrophysics, 357, 809

\bibitem[{Despali {et~al}\mbox{.}(2013)Despali, Tormen, \&
  Sheth}]{despali_ellipsoidal_2013}
Despali G., Tormen G., Sheth R.~K., 2013, Monthly Notices of the Royal
  Astronomical Society, 431, 1143

\bibitem[{Dodelson(2003)}]{dodelson}
Dodelson S., 2003, Modern Cosmology. Academic Press, Elsevier

\bibitem[{Doroshkevich(1970)}]{doroshkevich_spatial_1970}
Doroshkevich A.~G., 1970, Astrophysics, 6, 320

\bibitem[{Ehlers \& Buchert(1997)}]{ehlers_newtonian_1997}
Ehlers J., Buchert T., 1997, General Relativity and Gravitation, 29, 733

\bibitem[{Eisenstein \& Loeb(1995)}]{eisenstein_analytical_1995}
Eisenstein D.~J., Loeb A., 1995, The Astrophysical Journal, 439, 520

\bibitem[{Forero-Romero {et~al}\mbox{.}(2014)Forero-Romero, Contreras, \&
  Padilla}]{forero-romero_cosmic_2014}
Forero-Romero J.~E., Contreras S., Padilla N., 2014, Monthly Notices of the
  Royal Astronomical Society, 443, 1090

\bibitem[{Forero-Romero {et~al}\mbox{.}(2009)Forero-Romero, Hoffman,
  Gottlöber, Klypin, \& Yepes}]{forero-romero_dynamical_2009}
Forero-Romero J.~E., Hoffman Y., Gottlöber S., Klypin A., Yepes G., 2009,
  Monthly Notices of the Royal Astronomical Society, 396, 1815

\bibitem[{Fosalba \&
  Gaztanaga(1998{\natexlab{a}})}]{fosalba_cosmological_1998-1}
Fosalba P., Gaztanaga E., 1998{\natexlab{a}}, Monthly Notices of the Royal
  Astronomical Society, 301, 503

\bibitem[{Fosalba \& Gaztanaga(1998{\natexlab{b}})}]{fosalba_cosmological_1998}
Fosalba P., Gaztanaga E., 1998{\natexlab{b}}, Monthly Notices of the Royal
  Astronomical Society, 301, 535

\bibitem[{Gramann(1993)}]{gramann_second-order_1993}
Gramann M., 1993, The Astrophysical Journal Letters, 405, L47

\bibitem[{Hahn {et~al}\mbox{.}(2007)Hahn, Porciani, Carollo, \&
  Dekel}]{hahn_properties_2007}
Hahn O., Porciani C., Carollo C.~M., Dekel A., 2007, Monthly Notices of the
  Royal Astronomical Society, 375, 489

\bibitem[{Hoffman {et~al}\mbox{.}(2012)Hoffman, Metuki, Yepes, Gottl\"{o}ber,
  Forero-Romero, Libeskind, \& Knebe}]{hoffman_kinematic_2012}
Hoffman Y., Metuki O., Yepes G., Gottl\"{o}ber S., Forero-Romero J.~E.,
  Libeskind N.~I., Knebe A., 2012, Monthly Notices of the Royal Astronomical
  Society, 425, 2049, {arXiv}: 1201.3367

\bibitem[{Icke(1973)}]{icke_formation_1973}
Icke V., 1973, Astronomy and Astrophysics, 27, 1

\bibitem[{Johnson {et~al}\mbox{.}(2014)Johnson, Blake, Koda, Ma, Colless,
  Crocce, Davis, Jones, Magoulas, Lucey, Mould, Scrimgeour, \&
  Springob}]{johnson_6df_2014}
Johnson A. {et~al.}, 2014, Monthly Notices of the Royal Astronomical Society,
  444, 3926

\bibitem[{Joyce {et~al}\mbox{.}(2009)Joyce, Marcos, \&
  Baertschiger}]{joyce_towards_2009}
Joyce M., Marcos B., Baertschiger T., 2009, Monthly Notices of the Royal
  Astronomical Society, 394, 751

\bibitem[{Joyce \& Sylos~Labini(2012)}]{joyce_evolution_2012}
Joyce M., Sylos~Labini F., 2012, {ArXiv} e-prints, 1210, 1140

\bibitem[{Kayo {et~al}\mbox{.}(2001)Kayo, Taruya, \&
  Suto}]{kayo_probability_2001}
Kayo I., Taruya A., Suto Y., 2001, The Astrophysical Journal, 561, 22

\bibitem[{Kerscher {et~al}\mbox{.}(2001)Kerscher, Buchert, \& Futamase}]{kbt01}
Kerscher M., Buchert T., Futamase T., 2001, The Astrophysical Journal, 558, 79

\bibitem[{Kitaura {et~al}\mbox{.}(2012)Kitaura, Angulo, Hoffman, \&
  Gottlöber}]{kitaura_estimating_2012}
Kitaura F.-S., Angulo R.~E., Hoffman Y., Gottlöber S., 2012, Monthly Notices
  of the Royal Astronomical Society, 425, 2422

\bibitem[{Kitaura \& He{\ss}(2013)}]{kitaura_cosmological_2013}
Kitaura F.-S., He{\ss} S., 2013, Monthly Notices of the Royal Astronomical
  Society, 435, L78

\bibitem[{Kofman {et~al}\mbox{.}(1994)Kofman, Bertschinger, Gelb, Nusser, \&
  Dekel}]{kofman_evolution_1994}
Kofman L., Bertschinger E., Gelb J.~M., Nusser A., Dekel A., 1994, The
  Astrophysical Journal, 420, 44

\bibitem[{Kudlicki {et~al}\mbox{.}(2000)Kudlicki, Chodorowski, Plewa, \&
  Różyczka}]{kudlicki_reconstructing_2000}
Kudlicki A., Chodorowski M., Plewa T., Różyczka M., 2000, Monthly Notices of
  the Royal Astronomical Society, 316, 464

\bibitem[{Lam \& Sheth(2008{\natexlab{a}})}]{lam_ellipsoidal_2008}
Lam T.~Y., Sheth R.~K., 2008{\natexlab{a}}, Monthly Notices of the Royal
  Astronomical Society, 389, 1249

\bibitem[{Lam \& Sheth(2008{\natexlab{b}})}]{lam_perturbation_2008}
Lam T.~Y., Sheth R.~K., 2008{\natexlab{b}}, Monthly Notices of the Royal
  Astronomical Society, 386, 407

\bibitem[{Lee {et~al}\mbox{.}(2014)Lee, Rey, \& Kim}]{lee_alignments_2014}
Lee J., Rey S.~C., Kim S., 2014, The Astrophysical Journal, 791, 15

\bibitem[{Libeskind {et~al}\mbox{.}(2014)Libeskind, Hoffman, \&
  Gottl{\"o}ber}]{libeskind_velocity_2014}
Libeskind N.~I., Hoffman Y., Gottl{\"o}ber S., 2014, Monthly Notices of the
  Royal Astronomical Society, 441, 1974

\bibitem[{Lin {et~al}\mbox{.}(1965)Lin, Mestel, \&
  Shu}]{lin_gravitational_1965}
Lin C.~C., Mestel L., Shu F.~H., 1965, The Astrophysical Journal, 142, 1431

\bibitem[{Linder(2005)}]{linder_cosmic_2005}
Linder E.~V., 2005, Physical Review D, 72, 43529

\bibitem[{{\L}okas(2000)}]{lokas_universal_2000}
{\L}okas E.~L., 2000, Monthly Notices of the Royal Astronomical Society, 311,
  423

\bibitem[{Lynden-Bell(1964)}]{lynden-bell_large-scale_1964}
Lynden-Bell D., 1964, The Astrophysical Journal, 139, 1195

\bibitem[{Majerotto {et~al}\mbox{.}(2012)Majerotto, Guzzo, Samushia, Percival,
  Wang, de~la Torre, Garilli, Franzetti, Rossetti, Cimatti, Carbone, Roche, \&
  Zamorani}]{majerotto_probing_2012}
Majerotto E. {et~al.}, 2012, Monthly Notices of the Royal Astronomical Society,
  424, 1392

\bibitem[{Martino {et~al}\mbox{.}(2009)Martino, Stabenau, \&
  Sheth}]{martino_spherical_2009}
Martino M.~C., Stabenau H.~F., Sheth R.~K., 2009, Physical Review D, 79, 84013

\bibitem[{Matarrese \& Pietroni(2007)}]{matarrese_resumming_2007}
Matarrese S., Pietroni M., 2007, Journal of Cosmology and Astro-Particle
  Physics, 06, 026

\bibitem[{Matsubara(2008)}]{matsubara_resumming_2008}
Matsubara T., 2008, Physical Review D, 77, 63530

\bibitem[{Monaco(1995)}]{monaco_mass_1995}
Monaco P., 1995, The Astrophysical Journal, 447, 23

\bibitem[{Monaco {et~al}\mbox{.}(2013)Monaco, Sefusatti, Borgani, Crocce,
  Fosalba, Sheth, \& Theuns}]{monaco_accurate_2013}
Monaco P., Sefusatti E., Borgani S., Crocce M., Fosalba P., Sheth R.~K., Theuns
  T., 2013, Monthly Notices of the Royal Astronomical Society, 433, 2389

\bibitem[{Monaco {et~al}\mbox{.}(2002)Monaco, Theuns, \&
  Taffoni}]{monaco_pinocchio_2002}
Monaco P., Theuns T., Taffoni G., 2002, Monthly Notices of the Royal
  Astronomical Society, 331, 587

\bibitem[{Nadkarni-Ghosh(2013)}]{nadkarni-ghosh_non-linear_2013}
Nadkarni-Ghosh S., 2013, Monthly Notices of the Royal Astronomical Society,
  428, 1166

\bibitem[{Nadkarni-Ghosh \& Chernoff(2011)}]{nadkarni-ghosh_extending_2011}
Nadkarni-Ghosh S., Chernoff D.~F., 2011, Monthly Notices of the Royal
  Astronomical Society, 410, 1454

\bibitem[{Nadkarni-Ghosh \& Chernoff(2013)}]{nadkarni-ghosh_modelling_2013}
Nadkarni-Ghosh S., Chernoff D.~F., 2013, Monthly Notices of the Royal
  Astronomical Society, 431, 799

\bibitem[{Nadkarni-Ghosh \& Singhal(2015)}]{nadkarni-ghosh_phase_2015}
Nadkarni-Ghosh S., Singhal A., 2015, ArXiv e-prints, 1501, 7075

\bibitem[{Nariai \& Fujimoto(1972)}]{nariai_dynamics_1972}
Nariai H., Fujimoto M., 1972, Progress of Theoretical Physics, 47, 105

\bibitem[{Navarro {et~al}\mbox{.}(1996)Navarro, Frenk, \&
  White}]{navarro_structure_1996}
Navarro J.~F., Frenk C.~S., White S. D.~M., 1996, The Astrophysical Journal,
  462, 563

\bibitem[{Nusser {et~al}\mbox{.}(1991)Nusser, Dekel, Bertschinger, \&
  Blumenthal}]{nusser_cosmological_1991}
Nusser A., Dekel A., Bertschinger E., Blumenthal G.~R., 1991, The Astrophysical
  Journal, 379, 6

\bibitem[{Ohta {et~al}\mbox{.}(2003)Ohta, Kayo, \&
  Taruya}]{ohta_evolution_2003}
Ohta Y., Kayo I., Taruya A., 2003, The Astrophysical Journal, 589, 1

\bibitem[{Ohta {et~al}\mbox{.}(2004)Ohta, Kayo, \&
  Taruya}]{ohta_cosmological_2004}
Ohta Y., Kayo I., Taruya A., 2004, The Astrophysical Journal, 608, 647

\bibitem[{Peebles(1980)}]{peebles80}
Peebles P., 1980, The Large-Scale Structure of the Universe. Princeton
  University Press

\bibitem[{Press {et~al}\mbox{.}(2002)Press, Teukolsky, Vetterling, \&
  Flannery}]{numrecipes}
Press W., Teukolsky S., Vetterling W., Flannery B., 2002, Numerical Recipes in
  C++. Cambridge University Press

\bibitem[{Press \& Schechter(1974)}]{press_formation_1974}
Press W.~H., Schechter P., 1974, Astrophysical Journal, 187, 425

\bibitem[{Rossi(2012)}]{rossi_initial_2012}
Rossi G., 2012, Monthly Notices of the Royal Astronomical Society, 421, 296

\bibitem[{Scherrer \& Gaztañaga(2001)}]{scherrer_real-_2001}
Scherrer R.~J., Gaztañaga E., 2001, Monthly Notices of the Royal Astronomical
  Society, 328, 257

\bibitem[{Schneider {et~al}\mbox{.}(2012)Schneider, Frenk, \&
  Cole}]{schneider_shapes_2012}
Schneider M.~D., Frenk C.~S., Cole S., 2012, Journal of Cosmology and
  Astro-Particle Physics, 05, 030

\bibitem[{Sheth {et~al}\mbox{.}(2001)Sheth, Mo, \&
  Tormen}]{sheth_ellipsoidal_2001}
Sheth R.~K., Mo H.~J., Tormen G., 2001, Monthly Notices of the Royal
  Astronomical Society, 323, 1

\bibitem[{Stabenau \& Jain(2006)}]{stabenau_n-body_2006}
Stabenau H.~F., Jain B., 2006, Physical Review D, 74, 84007

\bibitem[{Susperregi \& Buchert(1997)}]{susperregi_cosmic_1997}
Susperregi M., Buchert T., 1997, Astronomy and Astrophysics, 323, 295

\bibitem[{White \& Silk(1979)}]{white_growth_1979}
White S. D.~M., Silk J., 1979, Astrophysical Journal, 231, 1

\bibitem[{Wintergerst \& Pettorino(2010)}]{wintergerst_clarifying_2010}
Wintergerst N., Pettorino V., 2010, Physical Review D, 82, 103516

\bibitem[{Zeldovich(1970)}]{zeldovich_gravitational_1970}
Zeldovich Y.~B., 1970, Astronomy and Astrophysics, 5, 84

\end{thebibliography}

\appendix
\section{Definitions of the tensors}
\label{app:tensors}
The physical and comoving axes of the ellipse are 
\beq
r_i = a_i q_i \; \; \; \; x_i = \frac{a_i}{a} q_i, 
\eeq
where $q_i$ is the initial Lagrangian coordinate of the $i$-th axis. Initially, the $a_i$ are different from $a$, so $q_i$ is not the initial comoving coordinate. If we want to characterize the deformation from the sphere then for the sphere $q_1=q_2 = q_3$. 
The deformation and velocity are 
\bea
s_i &=& q_i -x_i = \left(1-\frac{a_i}{a}\right) q_i \\
v_i &=& {\dot r}_i - H r_i =  \left(\frac{{\dot a}_i}{a_i} -H\right) r_i. 
\eea
The deformation tensor is 
\beq
e_{ij} = \frac{1}{2}\left( \frac{\partial s_i}{\partial q_j} + \frac{\partial s_j}{\partial q_i}\right) = \lambda_{a,i} \delta_{ij}.
\eeq
Note that we differ from BM96 by a minus sign. However, their relevant eigenvalues (denoted as $\lambda_{v,A}$) are negative of the eigenvalues of their $e_{ij}$. We have expressed it so that the $\lambda_{a}$ are eigenvalues of $e_{ij}$. This is just a matter of convention. The important point is that in both conventions $\lambda_a = \lambda_d$ in the linear regime and $\sum \lambda_{a,i} = \delta$. 
The tensor of velocity derivatives is 
\beq
\frac{1}{2H} \left( \frac{\partial v_i}{\partial r_j} + \frac{\partial v_j}{\partial r_i}\right) = \lambda_{v,i} \delta_{ij}. 
\eeq
The gravitational potential $\phi_p$ (scaled by $4 \pi G {\bar \rho_m}$) for the ellipsoid is 
\beq 
\phi_p = \frac{1}{2} \sum_i \lambda_{d,i} r_i^2.
\eeq
The Hessian of the gravitational potential is 
\beq 
\frac{\partial^2 \phi_p}{\partial r_i\partial r_j} = \lambda_{d,i} \delta_{ij}.
\eeq
The peculiar gravitational acceleration is ${\ddot r} = \nabla \phi_p$. Comparing with \eqnref{aieqn1} one can ensure that the definition is consistent.

\section{Derivation of the phase space equations}
\label{derivdyn}
 
With the definitions of the parameters the equation for the axes evolution takes the form 
\beq
\frac{{\ddot a}_i}{a_i} = -\frac{3H^2}{2}\left(\Omega_m\left\{\frac{1}{3} +\lambda_{d,i} \right\} -\frac{2}{3} \Omega_\Lambda\right). 
\label{eqnforai}
\eeq 
$\Omega_m$ and $\Omega_\Lambda$ are functions of $a$. 
\begin{enumerate}
\item Evolution of $\lambda_{a,i}$: from the definition $\lambda_{a,i} = 1-\frac{a_i}{a} $, it follows that 
\beq
{\dot \lambda}_{a,i} = -\left(\frac{{\dot a}_i}{a} - \frac{a_i}{a}H\right). 
\eeq
But from the definition of $\lambda_{v,i}$, ${\dot a}_i = H a_i (1+ \lambda_{v,i}) $. Substituting in the above equation, using the definition of $\lambda_{a,i}$ and converting from derivatives w.r.t. time to derivatives w.r.t. $\ln a$ gives 
\beq
\frac{ d\lambda_{a,i}}{d\ln a} = -\lambda_{v,i}(1-\lambda_{a,i}).  
\eeq

\item  Evolution of $\lambda_{v,i}$: from the definition $\lambda_{v,i} = \frac{1}{H}\frac{{\dot a}_i}{a_i} -1$, we have 
\beq
{\dot \lambda}_{v,i} = \frac{{\dot a}_i}{a_i} \left(1-\frac{1}{H^2} \frac{\ddot a}{a} \right) + \frac{1}{H} \left(\frac{{\ddot a}_i}{a_i} - \frac{{\dot a}_i^2}{a_i^2}\right).
\eeq
Using \eqnref{eqnforai}, the background evolution $\frac{{\ddot a}}{a} = -\frac{H^2}{2} (\Omega_m -2\Omega_\Lambda)$ and the definitions $\lambda_v$ gives 
\beq 
\frac{ d \lambda_{v,i}}{d \ln a} = -\frac{3}{2}\Omega_m\lambda_{d,i} +   \lambda_{v,i} \left(-1 + \frac{\Omega_m}{2} - \Omega_\Lambda \right) - \lambda_{v,i}^2.
\eeq
\item Evolution of $\lambda_{d,i}$: $\lambda_{d,i}$ is defined as 
\beq
\lambda_{d,i} = \frac{\delta \alpha_i}{2} + \frac{5}{4}\left(\alpha_i - \frac{2}{3}\right),
\label{lambdaddef}
\eeq
where
\bea
\delta &=& \frac{a^3}{a_1 a_2a_3}-1,\\
\alpha_i &=& a_1 a_2 a_3 f_i\\
{\rm and} \; \; \;   f_i &= &\int_0^\infty d \tau (a_i^2 + \tau)^{-\frac{3}{2}} \prod_{j=1}^3 (a_j^2 + \tau)^{-\frac{1}{2}}. 
\eea
Hence, 
\beq
\frac{d \lambda_{d,i}}{dt} = \left(\frac{\delta}{2}+\frac{5}{4}\right) \frac{d \alpha_i}{dt}  + \frac{\alpha_i}{2} \frac{ d\delta}{dt}. 
\label{eqn1}
\eeq
From the definitions, 
\bea
\label{eqn2}\frac{d \delta}{dt} &=& -H (1+ \delta) \sum_i\lambda_{v,i} \\
\label{eqn3}\frac{d \alpha_i}{dt} &=& H \alpha_i \sum_{i=1}^3 (1+ \lambda_{v,i}) + a_1 a_2 a_3 \frac{d f_i}{dt}.
\eea
The non-trivial part is the term $ \frac{d f_i}{dt}$. We proceed to evaluate it. Consider the case $i=1$.
\beq
\frac{d f_1}{dt} = 2 a_1 {\dot a_1} \int_{0}^\infty d\tau \left(-\frac{3}{2}(a_1^2 + \tau)^{-\frac{5}{2}}
(a_2^2 + \tau)^{-\frac{1}{2}}  (a_3^2 + \tau)^{-\frac{1}{2}} \right)  -a_2 {\dot a}_2 I_3 -a_3 {\dot a_3} I_2, 
\eeq
where 
\beq 
I_3 = \int_0^\infty d \tau (a_1^2 + \tau)^{-\frac{3}{2}}
(a_2^2 + \tau)^{-\frac{3}{2}}  (a_3^2 + \tau)^{-\frac{1}{2}} 
\eeq
and $I_2$ is obtained by interchanging $2$ and $3$ in $I_3$. Rewrite $-\frac{3}{2}(a_1^2 + \tau)^{-\frac{5}{2}}$ in the first integrand as $\frac{d (a_1^2 + \tau)^{-\frac{3}{2}}}{d \tau}$ and integrate by parts to give
\beq 
\frac{d f_1}{dt} = -\frac{2}{a_1 a_2 a_3} \frac{{\dot a}_1}{a_1} + (a_1 {\dot a}_1 -a_2 {\dot a}_2) I_3 + (a_1 {\dot a}_1 -a_2 {\dot a}_2) I_2
\eeq
The solutions for the integrals $I_2$ and $I_3$ can be found in a paper by Carlson (1987); equation 2.10. These expressions give the integrals $I_2$ and $I_3$ in terms of $R_F(a_1^2,a_2^2.a_3^2)$. 
\beq 
I_3 = \frac{1}{(a_1^2 - a_2^2)^2} \left(\frac{2}{3}(2 a_3^2 - a_1^2 - a_2^2) R_D(a_1^2, a_2^2, a_3^2) - 4 R_F(a_3^2, a_1^2, a_2^2) + \frac{a_1^2 + a_2^2}{a_1a_2 a_3} \right)
\eeq
But from relations in Carlson (1989), specifically, equations 2.12, 2.17 and 2.26, it can be shown that 
\beq
3 R_F(a_1^2,a_2^2,a_3^2) = a_1^2 R_D(a_2^2,a_3^2,a_1^2) + a_2^2 R_D(a_3^2,a_1^2,a_2^2) +a_3^2 R_D(a_1^2,a_2^2,a_3^2) 
\eeq
where 
\beq 
R_D(a_2^2,a_3^2,a_1^2) = \frac{3}{2} \frac{\alpha_1}{a_1 a_2 a_3}, \; \; \; R_D(a_3^2,a_1^2,a_2^2) = \frac{3}{2} \frac{\alpha_2}{a_1 a_2 a_3} \; \; \; {\rm etc}
\eeq
Using the two above results from Carlson's paper and the relation $\sum_i \alpha_i =2$, gives 
\beq 
a_1 a_2 a_3 \frac{d f_1}{dt} = -2\frac{{\dot a_1}}{a_1} + \frac{(\alpha_2 - \alpha_1) (a_1 {\dot a}_1 - a_2 {\dot a}_2)}{a_1^2 - a_2^2} + \frac{(\alpha_3 - \alpha_1) (a_1 {\dot a}_1 - a_3 {\dot a}_3)}{a_1^2 - a_3^2} 
\label{eqn4}
\eeq
Using \eqnrefs{eqn1}, \eqnrefbare{eqn2}, \eqnrefbare{eqn3} and \eqnrefbare{eqn4} along with the definitions of $\lambda_v$ and $\lambda_d$ gives
\bea
\frac{d \lambda_{d,i}}{d \ln a} &=& -(1+\delta) \left(\delta + \frac{5}{2} \right)^{-1} \left(\lambda_{d,i} + \frac{5}{6} \right)  \sum_{j=1}^3\lambda_{v,j} \\
\nonumber && + \left( \lambda_{d,i} + \frac{5}{6} \right) \sum_{i=1}^3 (1+\lambda_{v,i})  - \left( \delta + \frac{5}{2} \right)(1+\lambda_{v,i}) \\
\nonumber && + \sum_{j\neq i}\frac{ \left\{\lambda_{d,j} - \lambda_{d,i} \right\} \cdot \left\{(1-\lambda_{a,i})^2(1+\lambda_{v,i}) -(1-\lambda_{a,j})^2(1+\lambda_{v,j})\right\}}{(1-\lambda_{a,i})^2-(1-\lambda_{a,j})^2}.
\eea
\end{enumerate}
Note that only the equation for $\lambda_{v}$ depends on the background cosmology. The equations for $\lambda_a$ and $\lambda_d$ stay unchanged. 

{\it Spherical and linear limits:}
When all three axes are equal, $\lambda_d = \delta_{\mathrm{sph}}/3$ and $\lambda_v = \Theta_{\mathrm{sph}}/3$. It is easy to check that the $\lambda_v$ equation reduces to  
\eqnref{dthetasph} for the sphere. For case of the $\lambda_d$ equation, the last term can be shown to be tending to zero by substituting $\lambda_j = \lambda_i+\epsilon$ and taking the limit $\epsilon \rightarrow 0$. The second and third terms cancel and we are left with the first which reduces to \eqnref{ddeltasph}. 
Linearizing the $\lambda_v$ and $\lambda_d$ equations gives $d\lambda_{d,i}/d\ln a = - \lambda_{v,i}$ and $d\lambda_{v,i}/d\ln a =  \lambda_{v,i}$ giving the correct linear limit $ d\lambda_{v,i}/d\lambda_{d,i} = -1$. 

\section{Evolution of the signs of the $\lambda$ parameters}
\label{app:percentages}
\capfigref{percentages} shows the distribution of the signs of the $\lambda$ parameters. This distribution stays more or less constant throughout the evolution and may potentially change for initial conditions that are non-Gaussian. 

\begin{figure}
\includegraphics[width=16cm]{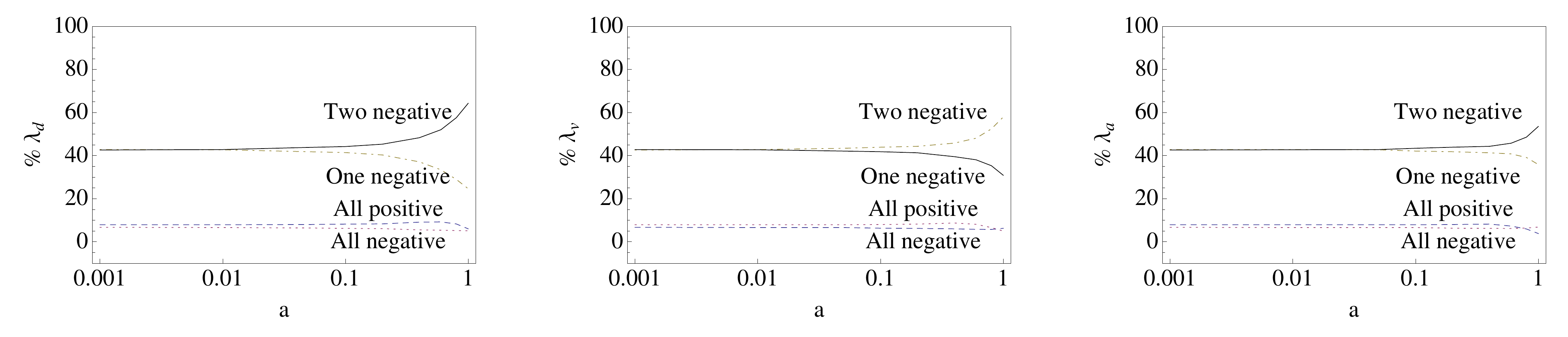}
\caption{The evolution of the signs of the individual $\lambda$ parameters. The initial pdf has a distribution of 42\% for two eigenvalues $\lambda_d$ positive or two negative. At early times this is the same for $\lambda_v$ and $\lambda_a$. We find that for all three $\lambda$ parameters, this distribution evolves in a similar manner changing only slightly at later epochs near $a=1$.}
\label{percentages}
\end{figure}

\section{Phase space dynamics for 1D collapse}
\label{1ddynamics}
If the perturbation is only along one axis and the other two axes are comoving with the background, then $\lambda_{d/v,2}=\lambda_{d/v,3}=0$. Then, $\lambda_{v,1} = \Theta$ and the diagonal components of the shear tensor are $\sigma_1= \frac{2}{3} \Theta$ and $\sigma_2= \sigma_3=-\frac{1}{3} \Theta$. Thus, the dynamical system given by \cref{ell_phasespace} for an EdS cosmology reduces to 
\begin{subequations}
\begin{align} 
\frac{d \delta}{d \ln a} &=-(1+ \delta) \Theta\\
\frac{d \Theta}{d \ln a} &= -\frac{1}{2}\left[3 \delta  + \Theta + 2 \Theta^2 \right], 
\end{align}
\end{subequations}
Rearrange the terms in the second equation (add and subtract $\delta$ and express $\Theta = 3 \Theta -2\Theta$) to give 
\beq
\frac{d \Theta}{d \ln a}=(1+\delta)\Theta -\frac{(\Theta + \delta) (3 + 2\Theta)}{2}.
\eeq
Dividing the second equation by the first, gives 
\beq
\frac{d \Theta}{d\delta} = -1 + \frac{(\Theta + \delta) (3 + 2\Theta) }{2 \Theta (1+ \delta)}
\eeq 
It is easy to check that a linear relation satisfies this equation. The initial conditions are $\Theta (\delta=0) =0$ and 
$\left.\frac{ d\Theta}{d\delta}\right|_{\delta=0} =c$, where $c$ is some pre-determined constant. Let $\Theta = A\delta + B$ be a solution to the above equation with the given initial conditions. The first initial condition sets $B=0$ and the second one gives 
\beq 
\left. \frac{ d\Theta}{d\delta}\right|_{\delta=0} = -1 + \frac{3}{2} (A+1) = c, 
 \eeq
which in turn sets $A = \frac{2}{3}(1+c) -1$. For $c=-1$, $A = -1$ giving $\Theta = -\delta$. 
A similar argument will hold for the $\Lambda$CDM case: only the coefficient of $\Theta$ changes so the rearrangement of terms will be slightly different and the constant $c$ (or alternatively $A$) will involve some factors of $\Omega_m$ and $\Omega_\Lambda$.

\section{Relation between the Lagrangian and Eulerian PDFs}
\label{EtoLframe}
 
Given a discrete distribution of massive particles with a mean density ${\bar \rho}$ there are two ways to define the density field. At each point, consider the amount of matter $M$ contained in a {\it fixed volume} $V_0$ and define the density contrast as $\delta(M, V_0)= \frac{1}{\bar \rho} \frac{M}{V_0}$. This corresponds to the Eulerian prescription. Alternately, in the Lagrangian prescription, at each point one estimates the volume occupied by a {\it fixed mass} $M_0$ and the corresponding density contrast is $\delta= \frac{1}{\bar \rho} \frac{M_0}{V}$. $M_0$ and $V_0$ are related through $\delta_0$ as 
\beq 
\delta_0 = \frac{1}{\bar \rho} \frac{M_0}{V_0} -1.  
\label{delta0M0}
\eeq
In the Eulerian prescription, $\delta$ and $M$ are equivalent variables whereas in the Lagrangian prescription, $\delta$ and $V$ are equivalent. Furthermore, a different choice of $V_0$ for the same choice of $M_0$ gives a different Eulerian density field and similarly a different choice of $M_0$ for a fixed choice of $V_0$ gives a different Lagrangian density field. To relate the PDFs of the Eulerian and Lagrangian fields, note that the probability that a given volume $V_0$ contains an amount of matter greater than $M_0$ is same as the probability that $M_0$ occupies a volume smaller than $V_0$ (B94 page 702, eq. 33a). This gives 
\beq 
\int_{M_0}^\infty p_E(V_0, M) dM = \int_0^{V_0} p_L(V,M_0) dV.  
\label{eul-lag0}
\eeq   
To change the integration variable to $\delta$ note that 
\bea
p_E(V_0, M) = p_E(V_0, \delta) \left|\frac{d\delta }{d M}\right| 
&\implies&  p_E(V_0, M) dM = p_E(V_0, \delta) d\delta \\
p_L(V,M_0) = p_L(\delta,M_0) \left|\frac{d \delta}{dV}\right| &\implies& p_L(V,M_0) dV = - p_L(\delta,M_0) d \delta.
\eea
The negative sign arises in the second equation because $\delta$ and $V$ are inversely related. The end points of the first intergal: $M = \{ M_0, \infty\} \implies \delta = \{\delta_0, \infty\}$ and in the second $V = \{0, V_0\} \implies\delta = \{\infty,\delta_0\}$. Thus, \eqnref{eul-lag0} becomes 
\beq 
\int_{\delta_0}^\infty p_E(V_0,\delta) d \delta = \int_{\delta_0}^\infty p_L(\delta,M_0) d\delta.  
\eeq  
Differentiating w.r.t. $\delta_0$ for a fixed $V_0$ gives 
\beq 
-p_E(V_0,\delta_0)  = - \left\{p_L(\delta_0,M_0) +  \int^{\delta_0}_\infty \frac{\partial}{\partial \delta_0} p_L(\delta,M_0) d\delta                                                                                                                                                             \right\}. 
\eeq
The above relation is between the discrete Eulerian and Lagrangian densities. 
Using the relation between the smooth and discrete densities given by B94 (i.e., the PDFs of the Eulerian densities differ by $1+\delta$, where as the PDFs of the Lagrangian densities are the same) gives 
\beq
(1+ \delta_0) p_E(V_0,\delta_0) = \left\{p_L(\delta_0,M_0) -  \int_{\delta_0}^\infty \frac{\partial}{\partial \delta_0} p_L(\delta,M_0) d\delta.
\label{eul-lag1}                                                                                                                                                              \right\}.
\eeq
The second term on the r.h.s. depends on the variation of $p_L$ with $\delta_0$ or alternately $M_0$.
 Ideally, to compute it we need to know how $p_L$ varies with $M_0$. A different mass scale corresponds to a different choice of $\sigma$. So we can rewrite 
 \bea
\frac{\partial p_L}{\partial \delta_0} & =& \frac{\partial p_L}{d M_0}\frac{\partial M_0}{d\delta_0}  \\
&=& \frac{\partial p_L}{d \sigma}\frac{\partial \sigma}{\partial M_0}  \frac{\partial M_0}{\partial \delta_0} \\
 &=&\frac{\partial p_L}{d \sigma}\frac{\partial \sigma}{\partial R_f}  \frac{\partial R_f}{\partial \delta_0},
  \eea
 where in the last equality we have introduced the filtering scale $R_f$ which corresponds to the mass scale $M_0$ through  
 \beq 
 M_0 = \frac{4}{3} \pi R_f^3 {\bar \rho}.
 \label{M0Rf}
 \eeq
 Combining the above with \eqnref{delta0M0},  
 \beq 
 (1+ \delta_0) V_0 = \frac{4}{3} \pi R_f^3
 \eeq
 and differentiating w.r.t. $\delta_0$ at a fixed $ V_0$ gives 
 \beq 
 \frac{d R_f }{d \delta_0} =  \frac{R_f}{3 (1+ \delta_0)}.
 \label{drfddelta0}
 \eeq
 Since $\frac{d \sigma}{d R_f}$ and  $\frac{d R_f}{d\delta_0}$ do not depend on $\delta$, \eqnref{eul-lag1} becomes 
 \beq
(1+ \delta_0) p_E(V_0,\delta_0) = \left\{p_L(\delta_0,\sigma) -  \frac{d \sigma}{d R_f}  \frac{d R_f}{d\delta_0} \int_{\delta_0}^\infty \frac{dp_L(\delta, \sigma) }{d\sigma}  d\delta.
\label{eul-lag2}                                                                                                                                                              \right\}.
\eeq
In \eqnref{eul-lag2}, $p_L$ corresponds to the non-linear probability distribution, however, the numerical studies in this paper were performed with only a few values of $\sigma$ and the dependence of the distribution on $\sigma$ cannot be known accurately. Instead, we examine the linear limit to obtain an analytic result and extend the result to the quasi-linear regime. This procedure can be partially justified by arguing that the theoretical results of B94 (the Eulerian pdf that we compare with) are expected to be valid only for small $\sigma$. 
 
In the linear regime, 
 \beq 
 p_L(\delta) = \frac{1}{\sqrt{2 \pi} \sigma} \exp{\left(-\frac{\delta^2}{2\sigma^2}\right)}. 
 \eeq
 \beq
 \frac{d p_L}{d\sigma} = -\frac{1}{\sqrt{2 \pi } \sigma^2} \exp{\left(-\frac{\delta^2}{2 \sigma^2}\right)} +\frac{ \delta^2}{\sqrt{2 \pi }  \sigma^4}\exp{\left(-\frac{\delta^2}{2 \sigma^2}\right)}.
 \eeq
 and the integral becomes 
 \beq 
  \int_{\delta_0}^\infty \frac{dp_L(\delta, \sigma) }{d\sigma}  d\delta = \frac{e^{-\frac{\delta_0^2}{2 \sigma^2}} \delta_0}{\sqrt{2 \pi } \sigma^2} = \frac{\delta_0}{\sigma} p_L. 
  \label{integral}
 \eeq
 It remains to compute $\frac{d\sigma}{d R_f}$ in \eqnref{eul-lag2}. This depends on the power spectrum and the window function. 
 \beq 
 \sigma^2(R_f) = \int P(k) W^2(k R_f) d^3 k 
 \eeq
 Assuming a Gaussian filter $W(kR_f) = \exp{-\frac{k^2 R_f^2}{2}}$ and a power spectrum with index $n$ i.e., $P(k) = c_0 k^n$, where $c_0$ is some constant, 
 \beq 
 \sigma^2 = \frac{c_0}{2} R_f^{-(n+3)} \Gamma\left(\frac{n+3}{2}\right); \; \; \; \; \; 2 \sigma \frac{d \sigma}{dR_f} = -\frac{n+3}{R_f} \sigma^2
 \eeq 
 and
\beq 
\frac{d \sigma}{dR_f} = -\frac{(n+3)}{2} \frac{ \sigma}{R_f}. 
\label{dsigmadrf}
\eeq
 Putting together \eqnrefs{drfddelta0}, \eqnrefbare{eul-lag2}, \eqnrefbare{integral} and \eqnrefbare{dsigmadrf}, gives 
\beq 
 (1+ \delta_0) p_E = p_L \left[1 + \frac{(n+3)}{6} \frac{\delta_0}{1+\delta_0}\right].
\eeq
In our comparisons, we used $n=-3$. In this case, $\sigma$ is independent of the smoothing scale and the relation is simply 
\beq 
p_E(\delta_0) = \frac{p_L(\delta_0)}{1+ \delta_0}. 
\eeq 
Arguing that the density-velocity divergence relation is the same in both frames (mean velocity divergence 
is computed from the velocity field (within a fixed volume in the Eulerian case and within a fixed mass in the Lagrangian case), the velocity PDFs also have the same relation: 
\beq 
p_E(\Theta) = \frac{p_L(\Theta)}{1+ \delta}. 
\eeq 
The Eulerian PDFs predicted by our analysis and the ones used for comparison are both normalized numerically. 

\end{document}